\documentclass[journal,comsoc]{IEEEtran}
\usepackage[T1]{fontenc}

\usepackage{algorithmic}
\usepackage{algorithm}
\usepackage{tabularx}
\usepackage{enumitem}
\setlist{leftmargin=5mm}
\ifCLASSOPTIONcompsoc
\usepackage[caption=false,font=footnotesize,labelfont=sf,textfont=sf]{subfig}
\else
\usepackage[caption=false,font=footnotesize]{subfig}
\fi
\usepackage{url}
\usepackage{graphicx}
\usepackage{amsmath}
\usepackage{amsfonts}
\usepackage{xcolor}
\usepackage{amssymb}
\usepackage{mathrsfs}
\usepackage{CJK}
\usepackage{textcomp}
\usepackage{epstopdf}
\ifCLASSOPTIONcompsoc
\usepackage[nocompress]{cite}
\else
\usepackage{cite}
\fi
\renewcommand{\algorithmicrequire}{\textbf{Input:}}
\renewcommand{\algorithmicensure}{\textbf{Output:}}
\renewcommand{\algorithmiccomment}[1]{\bgroup\hfill\scriptsize$\rhd$~#1\egroup}

\newcommand{\etal}{\emph{et al.}}
\newcommand{\R}{\mathcal{R}}

\ifCLASSINFOpdf
\else
\fi

\usepackage{amsmath}
\usepackage[cmintegrals]{newtxmath}
\begin{document}
	
\title{Fast Matrix Factorization with Non-Uniform Weights on Missing Data} 	
\author{Xiangnan~He,~
	Jinhui~Tang,~\IEEEmembership{Senior~Member,~IEEE,~}%
	Xiaoyu~Du,~
	Richang~Hong,~\IEEEmembership{Member,~IEEE,} 
	Tongwei~Ren,~\IEEEmembership{Member,~IEEE,}~%
	and~Tat-Seng~Chua
	
	\thanks{Xiangnan He is with the University of Science and Technology of China, Hefei, Anhui, China, 230031. E-mail: xiangnanhe@gmail.com}
	\thanks{Jinhui Tang is with the Nanjing University of Science and Technology, Nanjing, Jiangsu, China, 210094. E-mail: jinhuitang@njust.edu.cn. Corresponding author.}
	\thanks{Xiaoyu Du is with the University of Electronic Science and Technology of China, Chengdu, Sichuan, China, 610054. E-mail: duxy.me@gmail.com}
	\thanks{Richang Hong is with the Hefei University of Technology, Hefei, Anhui, China, 230000. E-mail: hongrc.hfut@gmail.com}%
	\thanks{Tongwei Ren is with the Nanjing University, Nanjing, Jiangsu, China, 210093. E-mail: rentw@nju.edu.cn} 
    \thanks{Tat-Seng Chua is with the National University of Singapore, Singapore, 117417. E-mail: dcscts@nus.edu.sg}
	}

\markboth{IEEE Transactions on Neural Networks and Learning Systems, Submission 2018}
{He \MakeLowercase{\textit{et al.}}: Fast Matrix Factorization with Non-Uniform Weights on Missing Data}

\maketitle

\begin{abstract}
	Matrix factorization (MF) has been widely used to discover the low-rank structure and to predict the missing entries of data matrix. In many real-world learning systems, the data matrix can be very high-dimensional but sparse.
    This poses an imbalanced learning problem, since the scale of missing entries is usually much larger than that of observed entries, but they cannot be ignored due to the valuable negative signal. For efficiency concern, existing work typically applies a uniform weight on missing entries to allow a fast learning algorithm. However, this simplification will decrease modeling fidelity, resulting in suboptimal performance for downstream applications. 
	
	In this work, we weight the missing data non-uniformly, and more generically, we allow any weighting strategy on the missing data. 
To address the efficiency challenge, we propose a fast learning method, for which the time complexity is determined by the number of observed entries in the data matrix, rather than the matrix size. The key idea is two-fold: 1) we apply truncated SVD on the weight matrix to get a more compact representation of the weights, and 2) we learn MF parameters with element-wise alternating least squares (eALS) and memorize the key intermediate variables to avoid repeating computations that are unnecessary. 
	We conduct extensive experiments on two recommendation benchmarks, demonstrating the correctness, efficiency, and effectiveness of our fast eALS method. 
\end{abstract}

\begin{IEEEkeywords}
Matrix Factorization, Missing Data, Element-wise Alternating Least Squares (eALS), Recommendation System.
\end{IEEEkeywords}

\IEEEpeerreviewmaketitle
\section{Introduction}\label{sec:introduction}
\IEEEPARstart
Matrices are a common data structure to represent the relation between two types of entities in learning systems~\cite{he2016fast,DBLP:pami/MaTLQZG15,Li:2016:EDL}. In relational learning, matrix factorization (MF) is a popular approach for dimension reduction by representing the rows (entities of one type) and columns (entities of another type) as two low-rank matrices. The optimization of dimension reduction is usually achieved by minimizing the reconstruction error between the low-rank model and the original data. Since the low-rank model can encode certain patterns latent in the original data, MF has been recognized as an effective pattern recognition technique and been widely used in a variety of tasks, such as relation prediction~\cite{TNNLS16Recom,DBLP:pami/TangSQLWYJ17}, data compression~\cite{de2017data}, clustering~\cite{He:WWW2014}, feature learning~\cite{ma2018tnnls}, topic modeling~\cite{li2017enhancing}, etc. 

When the relation of interest is sparse, it is always desired to predict whether a relation exists between two entities, known as relation prediction. The relation prediction task plays a vital role in many real-world learning systems, such as recommender systems~\cite{TNNLS16Recom,he2017neural}, language understanding~\cite{DBLP:ijcai/LeiWLI0K17,tang2016generalized}, social network mining~\cite{ZhaozhouSocial:TKDE,DBLP:tkde/LiaoLZ}, and so on. 
For relation prediction, which can be seen as a classification task, it is crucial to account for the missing entries, since they provide valuable signal about negative instances~\cite{yuan2018fbgd}. For example, in recommendation, early collaborative filtering approaches predict ratings by modeling the observed data only~\cite{Koren:SVD}; later on, researchers find that this way of ignoring missing data leads to poor performance in the real top-N recommendation system~\cite{Cremonesi:2010}. Another example is that in the learning of word embeddings~\cite{AllVec}, negative sampling is performed on missing data to add the negative signal about word-context co-occurrence, which is a crucial setup to ensure the semantics of learned word embeddings. 

Nevertheless, it is non-trivial to leverage the missing data, since its scale can be several orders of magnitude larger than the observed entries~\cite{yuan2018fbgd}. For example, in video recommendation data, users may only watch hundreds of videos on average among millions of videos, making the scale of missing data three orders of magnitude larger than the observed data in the user-video matrix~\cite{ACF}. This large-scale missing data poses efficiency challenges for learning the MF model. Towards this end, existing works have resorted to either sampling partial missing data as negative signal (aka., negative sampling ~\cite{Rendle:2009:BPR,yang2018did}) or modeling all missing data in a simplified way -- by assigning them a uniform weight to be negative~\cite{Hu:2008,devooght2015dynamic}. Both solutions have pros and cons: negative sampling has controllable efficiency, but its effectiveness may suffer from the low quality of negative examples and slow convergence~\cite{Rendle:2014:IPL,Ding:2018}; while modeling all missing data is costly, it can be more effective~\cite{he2016fast,AllVec}. To pursue high effectiveness, we focus on learning from all missing data in this work.

The Singular Value Decomposition (SVD)~\cite{klema1980singular} is a representative method for whole data-based MF. It assigns the same weight to all entries in the data matrix, regardless of whether they are observed or not (by default, the unobserved entries are assigned with a value of zero). This assumption makes the optimization problem have a nice structure and yields an analytical closed-form solution~\cite{srebro2003weighted}. However, considering that the number of missing data can be much larger than the observed data in real applications, it is more desirable to assign the missing data a lower weight to address the class imbalance issue. 
To this end, the Weighted Alternating Least Square (WALS)~\cite{Hu:2008} assigns a lower weight $c_0$ to all missing data, which is more flexible than the default setting of 1. 
However, we argue that WALS implicitly admits all missing data have the same likelihood to be negative, which may not be true in real applications. For example, in recommendation, we know that popular items are more likely to be known by users, and thus a missing on popular items is more likely to be a true negative. Lastly, it is worth mentioning that the uniform weight design in WALS is mainly due to the efficiency concern, since it allows for a clever speedup on ALS learning, which can avoid the high complexity brought by modeling all missing data. If we were to use non-uniform weights on missing data, the speedup trick of WALS is not applicable anymore, and the optimization complexity becomes unaffordable. 

In this work, we enhance 1) the flexibility of MF by allowing the use of non-uniform weights on missing data, and 2) the practicability of weighted MF by developing an efficient optimization algorithm. 
In short, we allow each missing entry to be assigned with an individualized weight, which encodes its prior to be a negative instance; the learning task takes the whole data matrix into account, but its time complexity is dependent on the number of observed entries only, rather than the matrix size (which is row\# $\times$ column\#). 
The two significant enhancements of our method make it easy to address large-scale relation prediction issue with a more expressive modeling on missing data, which has not been possible by the traditional MF methods like SVD and ALS. 
Our solution is achieved in three steps: 1) we perform truncated SVD on the weight matrix of missing data, using a more compact low-rank model to represent (or approximate) the weights of missing entries; 2) we perform ALS optimization on each element of user and item latent vectors, rather than the traditional vector-wise manner~\cite{srebro2003weighted,Hu:2008}; 3) we leverage the low-rank weights to design memoization strategies to reduce the time complexity significantly. 
Through comprehensive experiments on two real-world recommendation benchmarks, we verify the correctness and efficiency of our fast eALS method, and the effectiveness of using non-uniform weights on missing data for the recommendation task. 

A preliminary version of this work has been published as a conference paper in SIGIR 2016~\cite{he2016fast}. This paper is significantly different from its preliminary version in the methodology. Specifically, this work approaches a generic problem setting where any weighting strategy can be applied on missing data, but our previous work \cite{he2016fast} can only deal with a simpler case where the missing entries of a column have the same weight; moreover, the recent work \cite{li2018cf} can also be seen as a simpler case of this work where the missing entries of a row have the same weight.   
As such, the Preliminaries (Section~\ref{sec:preliminary}), Proposed Methods (Section~\ref{sec:method}), and Experiments (Section~\ref{sec:experiments}) have been re-written to support our solution to the new generic problem. 
The key contributions of this paper are summarized as follows:
\begin{itemize}
	\item We highlight the problem of optimizing MF with non-uniform weights on missing data and present an element-wise ALS algorithm to solve it. 
	\item We propose a fast eALS algorithm that solves the weighted MF problem with low-rank weights on missing data. The algorithm has a low time complexity in proportion to the number of observed entries and is independent of the number of missing entries. 
	\item We perform extensive experiments on two real-world datasets and demonstrate its correctness, efficiency, and effectiveness. The codes of our experiments can be found in: \url{https://github.com/duxy-me/ext-als}. 

\end{itemize}


\section{Preliminaries}\label{sec:preliminary}
This section provides some preliminaries about MF and formalizes the problem to solve in this paper. Moreover, we discuss the efficiency challenge in solving the problem. Note that part A of this section has been presented in the preliminary version \cite{he2016fast} (cf. Section 3.1) and other parts are new.  
Before starting the section, we first introduce some notations. 

We denote the original data matrix as $\textbf{R}\in \mathbb{R}^{M\times N}$, where $M$ and $N$ denote the number of rows and columns in the data matrix, respectively. We use the set $\R$ to denote the set of observed entries in $\textbf{R}$, i.e., for which the values are non-zero. Matrices $\textbf{P}\in \mathbb{R}^{M\times K}$ and $\textbf{Q}\in \mathbb{R}^{N\times K}$ denote the latent factor matrix for rows and columns respectively; that is, they are the results or model parameters of MF. 
We use the vector $\textbf{p}_u$ to denote the $u$-th row of matrix $\textbf{P}$, and we use the set $\R_u$ to denote the column indices with a nonzero value on row $u$, i.e., $\R_u = \{i | r_{ui} \neq 0\}$. We use the symbols $\textbf{q}_i$ and $\R_j$ to denote the similar meanings for the column side. Throughout the paper, we use the uppercase bold font to denote a matrix, lowercase bold font to denote a vector, and lowercase italic font to denote a scalar; for example, $\textbf{P}$ denotes a matrix, $\textbf{p}_u$ denotes the $u$-th row vector in $\textbf{P}$, and $p_{ui}$ denotes the $(u,i)$-th entry in $\textbf{P}$.

\subsection{The MF model}\label{subsec:mfmodel}

MF maps both rows and columns into a low-dimension latent space~(the dimension is $K$) such that their interactions are modeled as an inner product in that space~\cite{Koren:SVD}. Mathematically, each element $r_{ui}$ of $\textbf{R}$ is estimated as:
\begin{equation}
\label{eq:mf}
\hat{r}_{ui} = <\textbf{p}_u, \textbf{q}_i> = \textbf{p}_u^T \textbf{q}_i,
\end{equation}
where $\textbf{p}_u$ and $\textbf{q}_i$ are model parameters, which can be understood as the latent feature vector for row $u$ and column $i$, respectively. The model estimation $\hat{r}_{ui}$ can be seen as reconstruction for an observed entry, or prediction for an unobserved one. For example, in recommendation, $r_{ui}$ denotes a user's rating on an item (the larger the better), and ranking all items by $\hat{r}_{ui}$ can be used to select top-N recommendations for $u$. In matrix-wise representation, the model can be expressed as $\textbf{R} \approx \textbf{P}\textbf{Q}^T$, which implies the low-rank assumption of the data matrix~\cite{EFM}. 

%

\subsection{Problem Formulation}
Since MF performs dimension reduction on the original data matrix ($K$ is typically set to be much smaller than $M$ and $N$), the objective function for model learning is usually formed as an error-based regression loss~\cite{Koren:SVD,srebro2003weighted}. In this work, we learn MF parameters by solving the minimization problem on the objective function as follows:
\begin{equation}
\label{eq:pwl}
\begin{aligned}
L &= \sum_{u=1}^M\sum_{i=1}^N w_{ui}(r_{ui} - \textbf{p}_u^T\textbf{q}_j)^2 + \lambda(\sum_{u=1}^M||\textbf{p}_u||^2 + \sum_{i=1}^N||\textbf{q}_i||^2) \\
&= ||\textbf{W}\odot(\textbf{R} - \textbf{P}\textbf{Q}^T) ||^2 + \lambda(||\textbf{P}||^2 + ||\textbf{Q}||^2),
\end{aligned}
\end{equation}
where $w_{ui}$ denotes the weight of the training instance $r_{ui}$, $\textbf{W}\in \mathbb{R}^{M\times N}$ is the matrix form for all weights $w_{ui}$, and $\lambda$ is a hyper-parameter to control the regularization strength to prevent overfitting. In this problem formulation, we consider all data entries in $\textbf{R}$ and assign each data entry with an individualized weight $w_{ui}$, which is a generic setting that gives practitioners the flexibility to design the weighting strategy. Many previous efforts on MF do not deal with this generic problem setting, but instead use a specific weighting strategy. Here we discuss three most common strategies:

\textbf{Strategy 1. Zero weight on missing entries.} This strategy applies a zero weight on missing entries, i.e., $w_{ui}=0\  \text{if}\ (u,i) \notin \R$. Since only observed entries are used as training instances, the learning time complexity is low, which depends on the number of observed entries. 
This is a typical setting for the rating prediction task~\cite{Koren:SVD,DBLP:tkde/WangTW018}, which aims to predict the values of missing entries in user-item rating matrix. When the data follows the missing at random (MAR) assumption, such a setting can provide unbiased estimation. However, the MAR assumption does not hold in many real-world applications, for example, a user is are more likely to rate movies of her interest, rather than a random set of movies~\cite{Marlin07MAR}. In this case, the missing entries contain valuable signal about negative instances, and thus ignoring them will lead to suboptimal performance, especially for predicting whether a relation exists between two entities~\cite{Cremonesi:2010}. 
  
\textbf{Strategy 2. Uniform weight on all entries.} This strategy applies a uniform weight of 1 on all data entries, i.e., $w_{ui}=1\ \text{for all}\ (u,i)$. The SVD method~\cite{klema1980singular} can be directly applied to find the optimal solution for this problem. When the number of missing entries are of the same scale as the number of observed entries, such a setting may yield good performance. However, many real-world applications need to deal with sparse matrix that is highly imbalanced, for example, the observed ratings only take $1.2\%$ of the rating matrix in the Netflix challenge data\footnote{\url{https://en.wikipedia.org/wiki/Netflix_Prize}}. For such highly imbalanced learning scenarios, a uniform weighting strategy will make the parameter estimation process dominated by the missing entries, resulting in suboptimal performance. 

\textbf{Strategy 3. Uniform weight on missing entries.} This strategy assigns all missing entries with the same weight $c_0$, which can be different as the weight for observed entries~\cite{Hu:2008}:
\begin{equation} 
\begin{aligned}
w_{ui} = \begin{cases}
c_{ui} & if (u,i) \in \R, \\
c_0 & if (u,i) \notin \R, 
\end{cases}
\end{aligned}
\end{equation}
where $c_{ui}$ denotes the weight of observed entry $(u,i)$, and $c_0$ denotes the uniform weight for all missing entries. When dealing with sparse data, $c_0$ can be set as a smaller number than $c_{ui}$ to alleviate the imbalanced learning issue. Hu et al.~\cite{Hu:2008} demonstrated that this strategy yields better performance than a uniform weight on all entries in recommendation task. 
However, the deficiency is that it assumes all missing entries provide the same level of negative signal, which severely limits the fidelity for modeling real-world scenarios. For example, in recommendation systems, we know that the exposed but unclicked items (e.g., display ads) are more likely to be true negatives~\cite{Ding:2018}, which should be assigned with a higher weight than others. Another reasonable intuition is that the missing entries of active users (who have consumed many items) are more likely to be true negatives~\cite{li2018cf}.

\noindent In this work, we do not assume any weighting strategy on the data entries, and provide a solution for solving the generic problem of Eq.~(\ref{eq:pwl}). In other words, our solution subsumes the above-mentioned works that define various weighting strategies. 

\subsection{Efficiency Discussion}
One key reason that the previous work assumes a specific weighting strategy for missing data is due to efficiency concern. Here we analyze the learning time complexity for the three strategies: 

- For Strategy 1, the training set contains only $|\R|$ observed entries, thus standard optimization method like stochastic gradient descent~(SGD) can be applied, which has the time complexity of $O(|R|K)$. This level of complexity is rather low, only requiring a traversal on all training instances and updating latent vectors $\textbf{p}_u$ and $\textbf{q}_i$ at each visit of instance $(u,i)$. 

- For Strategy 2, since SVD can be directly applied to find the global optimum solution, the learning complexity depends on the solver for SVD. The commonly used solver Lanczos Bidiagonalization~(LBD) method \cite{komzsik2003lanczos} has a complexity linear with respect to the number of observed entries.
As such, its actual running time is in the same magnitude as the SGD solver for Strategy 1, and we can denote the analytical time complexity of SVD as $O(|R|K)$. 

- For Strategy 3, the training set contains all $M\times N$ data entries, for which standard optimization methods like SGD have the time complexity of $O(MNK)$. This level of complexity is rather high, being unaffordable for real-world large-scale applications that may have over millions of rows and columns (e.g., the user-item matrix in recommendation). Fortunately, the uniform weight constraint brings opportunities for speedup by memorizing some intermediate variables. Hu et al. \cite{Hu:2008} leveraged on memoization tricks and proposed an ALS-based algorithm (named as WALS), reducing the time complexity to $O(|\R|K^2 + (M+N)K^3)$.
Note that the $O(K^3)$ term is brought by the matrix inversion operation, which is inevitable when optimizing a latent vector (i.e., $\textbf{p}_u$ or $\textbf{q}_i$) as a whole in ALS~\cite{he2016fast}.  
Even so, the $O(|\R|K^2)$ part is still more costly than SGD, which only requires $O(|\R|K)$ time. 
As a result, even with speedup, WALS may still be prohibitive for running on large data, where large $K$ is crucial as it can lead to better representation ability and better performance. Lastly, it is worth mentioning that the speedup design in WALS is only applicable when the missing entries have the same weight. When such a nice structure is broken, WALS degrades to ALS with a complexity of $O(MNK)$. 

In this paper, we propose a new solution to efficiently solve the weighted MF problem. Distinct from the above-mentioned efforts that performed vector-wise (i.e., $\textbf{p}_u$) or matrix-wise (i.e., \textbf{P}) optimization, we perform optimization on the element level (i.e., $p_{uk}$), named as \textit{element-wise} ALS (or eALS for short). Furthermore, we apply a low-rank model to represent the weights of missing entries. Unifying the two designs, our solution achieves a time complexity of the $O(|R|K)$ level, but is more flexible on the weights of missing entries. 

\section{Proposed Methods}\label{sec:method}

We first present a vanilla element-wise ALS learner, which differs from the conventional vector-wise ALS~\cite{srebro2003weighted,Hu:2008}. 
By performing optimization on each element of the parameter matrix $\textbf{P}$ and $\textbf{Q}$, not only we can avoid the expensive matrix inversion operation in optimization, but also allow for more flexible design of memoization strategies for further speedup. 
Next, we propose to represent the weights for missing entries with a low-rank model, which not only reduces the space to store the weights for missing data, but also opens the door for speeding up the eALS learner. Lastly, based on the low-rank weights, we elaborate the fast eALS algorithm and discuss its several properties. Note that part A of this section has presented in the preliminary version \cite{he2016fast} (cf. Section 3.3) and other parts are different. 

\subsection{Vanilla Element-wise ALS Learner}
\label{ss:generic_eals}
One bottleneck of the previous WALS solution lies in the matrix inversion operation, which is due to the updating of the latent vector for a row (column) as a whole~(more explanations see Section 3.2 of \cite{he2016fast}). As such, it is a natural thought to avoid this operation by optimizing parameters at the element level.
Specifically, we follow the coordinate descent setting~\cite{libfm,iCD} that optimizes each element of the latent vector while leaving the others fixed.

First, we differentiate the objective function Eq.~(\ref{eq:pwl}) with respect to $p_{uf}$:
\begin{equation}\label{eq:d_puf}
\frac{\partial J}{\partial p_{uf}} = -2\sum_{i=1}^N (r_{ui} - \hat{r}_{ui}^f)w_{ui}q_{if} + 2p_{uf}\sum_{i=1}^N w_{ui}q_{if}^2 + 2\lambda p_{uf},
\end{equation}
where $\hat{r}_{ui}^f = \hat{r}_{ui} - p_{uf}q_{if}$, which can be understood as the model prediction in the absence of latent factor $f$. 
Given other variables fixed, the optimal solution of  $p_{uf}$ can be obtained at the point of $\frac{\partial J}{\partial p_{uf}} = 0$. Solving this equation, we can have:
\begin{equation}
\label{eq:p_uf0}
p_{uf} = \frac{\sum_{i=1}^N (r_{ui} - \hat{r}_{ui}^f)w_{ui}q_{if}}{\sum_{i=1}^N w_{ui}q_{if}^2 + \lambda}.
\end{equation}

Following the similar way, we can get the solver for item latent factor $q_{if}$:
\begin{equation}
\label{eq:q_if0}
q_{if} = \frac{\sum_{u=1}^M (r_{ui} - \hat{r}_{ui}^f)w_{ui}p_{uf}}{\sum_{i=1}^M w_{ui}p_{uf}^2 + \lambda}.
\end{equation}

With the above solution that solves on variable with others fixed, we can get a learning algorithm  by iteratively executing on all parameters until convergence. It is worth noting that since the objective function is non-convex in terms of all parameters together. 
As such, this element-wise ALS solver can only find local minima (where the critical points where gradients vanish). This is the same for the conventional ALS \cite{Hu:2008,Pan:2008} and other gradient descent methods in optimizing the objective function. As a consequence, the initialization of model parameters will affect the results. According to our experiment experience, eALS's performance is relatively stable with a Gaussian random initialization. 

\begin{algorithm}
	\caption{Vanilla eALS algorithm for weighted MF.}
	\label{alg:gefm}
	\begin{algorithmic}[1]
		\REQUIRE{Data matrix \textbf{R}, weight matrix \textbf{W}, number of latent factors $K$,  regularizer $\lambda$}
		\ENSURE{MF parameters \textbf{P} and \textbf{Q}}
		\STATE Randomly initialize $\textbf{P}$ and $\textbf{Q}$;
		\FOR{$(u,i)\in \R$}
		\STATE $\hat{r}_{ui} \leftarrow$ Eq. (\ref{eq:mf});
		\ENDFOR
		\WHILE{Stopping criteria is not met} 
		\vspace{1mm}
		\STATE // Update $\textbf{P}$  \COMMENT{$O(NMK)$}
		\FOR {{$u \leftarrow 1$ \textbf{to} $M$}}
		\FOR {{$f \leftarrow 1$ \textbf{to} $K$}}
		\STATE \textbf{for} $i \leftarrow 1$ \textbf{to} $N$ \textbf{do}  $\hat{r}_{ui}^f\leftarrow \hat{r}_{ui} - p_{uf}q_{if}$;
		\STATE $p_{uf}\leftarrow$ Eq.~(\ref{eq:p_uf0})  \COMMENT{$O(N)$}
		\STATE \textbf{for} $i \leftarrow 1$ \textbf{to} $N$ \textbf{do}  $\hat{r}_{ui}\leftarrow \hat{r}_{ui}^f + p_{uf}q_{if}$;
		\ENDFOR
		\ENDFOR
		\vspace{1mm}
		\STATE // Update $\textbf{Q}$ \COMMENT{$O(NMK)$}
		\FOR {{$i \leftarrow 1$ \textbf{to} $N$}}
		\FOR {{$f \leftarrow 1$ \textbf{to} $K$}} 
		\STATE \textbf{for} $u\leftarrow 1$ \textbf{to} $M$ \textbf{do}  $\hat{r}_{ui}^f\leftarrow \hat{r}_{ui} - p_{uf}q_{if}$\;
		\STATE $q_{if}\leftarrow$ Eq.~(\ref{eq:q_if0})  \COMMENT{$O(M)$} 
		
		\STATE \textbf{for} $u\leftarrow 1$ \textbf{to} $M$ \textbf{do} $\hat{r}_{ui}\leftarrow \hat{r}_{ui}^f + p_{uf}q_{if}$\;
		\ENDFOR
		\ENDFOR
		\ENDWHILE
	\end{algorithmic}
\end{algorithm}

\textbf{Time Complexity}.
We can see that by updating each element $p_{uf}$ and $q_{if}$ at a time, we can avoid the expensive matrix inversion operation, which is compulsory in traditional ALS. Through this way, we can eliminate the $O((M+N)K^3)$ term in the time complexity. 
Furthermore, we can follow the caching strategy as stated in \cite{libfm}, further reducing the time complexity from $O(M N K^2)$ (i.e., the time complexity of directly implementing the update rules) to $O(M N K)$. This time complexity is in the same level as evaluating all points in the data matrix $\textbf{R}$. Algorithm~\ref{alg:gefm} shows the vanilla eALS algorithm with the cache on $\hat{r}_{ui}$, which has a time complexity of $O(MNK)$. 

\subsection{Low-Rank Representation on Weights of Missing Data}
\label{ss:low_rank_weights}
The original problem of weighted MF assigns an individualized weight for each data point, which requires $O(MN)$ space to store all weights (denoted as the weight matrix $\textbf{W}$). 
This is very costly and unrealistic for large-scale applications. To be more specific, let us consider an intuitive example in recommendation that needs to deal with 1 million users and 1 million items. Assuming we use the 4-Byte float type to express a weight, then the space to store all weights is $1e^6\times1e^6\times4B = 4000 GB = 4TB$. Such a large consumption of space will pose a great challenge for the infrastructure, not to mention that the space cost will increase quadratically with respect to the number of users and items. 

Now that storing an individualized weight for each data point is practically infeasible, we consider using a more compact way to represent $\textbf{W}$. Specifically, we perform truncated SVD on $\textbf{W}$, obtaining two low-rank matrices which can reconstruct $\textbf{W}$ without any error:
\begin{equation}
\textbf{W} = \textbf{A} \textbf{B}^T
\end{equation}
where $\textbf{A}\in \mathbb{R}^{M\times Z'}, \textbf{B} \in \mathbb{R}^{N\times Z'}$, and $Z'$ denotes the rank size of weight matrix $\textbf{W}$. If $Z'$ is much smaller than $M$ and $N$, using $\textbf{A}$ and $\textbf{B}$ to reconstruct $\textbf{W}$ takes fewer space $O((M+N)Z'))$ than directly storing $\textbf{W}$. If $Z'$ is large such that the space complexity of $O((M+N)Z'))$ is still unaffordable, one can use truncated SVD with a smaller number of predefined rank size, but at the cost of approximating $\textbf{W}$ with some errors --- the smaller the number, the larger the error of the approximation. This is a tradeoff between the space cost and the precision of low-rank representation. To be precise, let $Z$ be the predefined rank size of truncated SVD, then the space cost to store the weights is $O((M+N)Z))$. Since the common solver of SVD like the Lanczos Bidiagonalization~(LBD) method \cite{komzsik2003lanczos} has a complexity linear with respect to the number of observed entries, the analytical time complexity of truncated SVD is $O(|\mathcal{R}|Z)$. Given $Z$ is usually a small number (e.g., 1 for column-oriented~\cite{he2016fast} or row-oriented~\cite{li2018cf} weighting schemes), this time complexity is much lower than matrix factorization algorithms (which are shown in Table 1). As such, using truncated SVD on the weights will not significantly increase the actual runtime of our method. 

In a sparse matrix, the number of missing entries is usually several magnitudes than the number of observed entries. As such, we apply truncated SVD on the weights of missing entries only, and use the original weights of observed entries as they are. Such a weighting strategy can be expressed as follows:
\begin{equation} \label{eq:wui}
\begin{aligned}
w_{ui} = \begin{cases}
c_{ui} & \ \text{if} (u,i) \in \R, \\
\textbf{a}_u^T\textbf{b}_i & \  \text{if} (u,i) \notin \R, 
\end{cases}
\end{aligned}
\end{equation}
where $c_{ui}$ denotes the weight of observed entry (u,i), $\textbf{a}_u \in \mathbb{R}^{Z}$ and $\textbf{b}_i\in \mathbb{R}^{Z}$ denote the $u$-th row vector of $\textbf{A}$ and $i$-th row vector of $\textbf{B}$, respectively. In the next subsection, we present a fast algorithm to accelerate eALS by leveraging the low-rank structure of the weights of missing data. 



\subsection{Fast eALS Algorithm}
\label{ss:fast_eals}
The fast algorithm to be presented in this part reduces the time complexity from $O(MN)$-related to $O(|\R|)$-related, successfully avoiding the heavy burden brought by optimizing the missing data. 

First, we reformulate the objective function by separating the terms on observed data and missing data:
\begin{equation} \label{eq:partitioned}
\begin{aligned}
J &= \sum_{(u,i) \in \mathcal{R}}c_{ui}(r_{ui} - \hat{r}_{ui})^2
+ \sum_{(u,i) \notin \mathcal{R}}\textbf{a}_u^T\textbf{b}_i(r_{ui} - \hat{r}_{ui})^2\\
&+ \lambda(\sum_{u=1}^M||\textbf{p}_u||^2 + \sum_{i=1}^N||\textbf{q}_i||^2)
\end{aligned}
\end{equation}
As we can see, the first term focuses on the observed data only and leads to a low complexity in optimization. The major cost comes from the second term that operates on all missing data. Next, we elaborate the derivation process of optimizing user latent factor $p_{uf}$, and its counterpart of optimizing item latent factor $q_{if}$ can be achieved similarly. \vspace{+10pt}

\noindent First, we compute the derivative of $J$ with respect to $p_{uf}$ and set the derivative to zero, we can obtain the update rule of $p_{uf}$:
\begin{equation}\label{eq:p_uf}
p_{uf} = \frac{\sum_{i\in \mathcal{R}_u} (r_{ui} - \hat{r}_{ui}^f)c_{ui}q_{if} - \sum_{i\notin \mathcal{R}_u}  \textbf{a}_u^T \textbf{b}_i\hat{r}_{ui}^fq_{if} }
{\sum_{i\in \mathcal{R}_u} c_{ui}q_{if}^2 + \sum_{i\notin \mathcal{R}_u} \textbf{a}_u^T\textbf{b}_i q_{if}^2 + \lambda}.
\end{equation}
where $\hat{r}_{ui}^f = \hat{r}_{ui} - p_{uf}q_{if}$, i.e., the prediction without the component of latent factor $f$. 
With careful inspection, we can find that
the major cost of executing the update rule comes from the two sum operations on missing data, i.e., $\sum_{i\notin \mathcal{R}_u}\textbf{a}_u^T \textbf{b}_i\hat{r}_{ui}^fq_{if}$ (in the numerator) and $\sum_{i\notin \mathcal{R}_u} \textbf{a}_u^T\textbf{b}_i q_{if}^2$ (in the denominator). 
We term the evaluation of the two costly terms as the $rq$ problem and the $q^2$ problem, respectively, and show how to solve the two problems in an efficient way. 

\textbf{1. Solving the $rq$ problem}. First, we expand the term $\textbf{a}_q^T\textbf{b}_i$ using element-wise operations and obtain:
\begin{equation}
	\sum_{i\notin \R_u}\textbf{a}_u^T\textbf{b}_i\hat{r}_{ui}^fq_{if} = \sum_{i\notin \R_u}\sum_{t=1}^Za_{ut}b_{it}\sum_{k \neq f}p_{uk}q_{ik}q_{if}.
\end{equation}
Then, we re-arrange the sum operations and obtain its equivalent form:
\begin{equation}\small
\begin{aligned}
	\sum_{i\notin \R_u}\textbf{a}_u^T\textbf{b}_i\hat{r}_{ui}^fq_{if} &= \sum_{k\neq f}p_{uk}\sum_{t=1}^Za_{ut}\sum_{i\notin \R_u}b_{it}q_{ik}q_{if} \\
	&= \sum_{k\neq f}p_{uk}\sum_{t=1}^Za_{ut}(\sum_{i=1}^Nb_{it}q_{ik}q_{if} - \sum_{i\in \R_u}b_{it}q_{ik}q_{if})
\end{aligned}
\end{equation}
\noindent As we can see, the main computational bottleneck is in the term $\sum_{i=1}^Nb_{it}q_{ik}q_{if}$, which needs to scan over all items. Nevertheless, a nice property is that this term is independent of $u$ --- which means if we sequentially update all elements in $\textbf{P}$, and then $\textbf{Q}$, this term can be pre-computed and used for the updating of all elements in $\textbf{P}$ without computing it on-the-fly (the reverse way also applies). 
To achieve this, we define a 3-dimensional tensor $\textbf{S}^q \in \mathbb{R}^{T\times F\times K}$, in which each element is defined as $S_{tfk}^q = \sum_{i=1}^Nb_{it}q_{ik}q_{if}$; the $\textbf{S}^q$ tensor is computed after updating $\textbf{Q}$ and is cached in the updates of $\textbf{P}$. 
With this cache, the $rq$ problem can be approached as:
\begin{equation} \small
\begin{aligned}
\sum_{i\notin\R_u}\textbf{a}_u^T\textbf{b}_i\hat{r}_{ui}^fq_{if} &= \sum_{k\neq f}p_{uk}\sum_{t=1}^Z a_{ut}(S_{tkf}^q - \sum_{i\in \R_u}b_{it}q_{ik}q_{if}),
\end{aligned}
\end{equation}
which can be computed in $O(KZ|\R_u|)$ time, rather than the raw time complexity of $O(KZMN)$. 

\textbf{2. Solving the $q^2$ problem}. We can apply the similar cache strategy to address the costly $q^2$ problem:
\begin{equation}
\begin{aligned}
\sum_{i\notin \R_u} \textbf{a}_u^T\textbf{b}_i q_{if}^2
&= \sum_{i\notin \R_u}\sum_{t=1}^Z a_{ut}b_{it}q_{if}^2\\
&= \sum_{t=1}^Z a_{ut}(\sum_{i=1}^Nb_{it}q_{if}^2 - \sum_{i\in \R_u}b_{it}q_{if}^2)\\
&= \sum_{t=1}^Z a_{ut}S_{tff}^q - \sum_{i\in \R_u}\textbf{a}_u^T\textbf{b}_iq_{if}^2,
\end{aligned}
\end{equation}
where $S_{tff}^q$ denotes the $(t,f,f)$-th element of the $\textbf{S}^q$ cache. \vspace{+10pt}


\noindent We can apply the similar derivation process on the item latent factor $q_{if}$ to obtain its update rule:
\begin{equation}\label{eq:q_if}
	q_{if} = \frac{\sum_{u\in \R_i} c_{ui}(r_{ui} - \hat{r}_{ui}^f)p_{uf} - \sum_{u\notin \R_i} \textbf{a}_u^T\textbf{b}_i \hat{r}_{ui}^fp_{uf} }{\sum_{u\in \R_i}c_{ui}p_{uf}^2+\sum_{u\notin \R_i}\textbf{a}_u^T\textbf{b}_ip_{uf}^2+\lambda},
\end{equation}
To speed up the computation of $\sum_{u\notin \R_i} \textbf{a}_u^T\textbf{b}_i \hat{r}_{ui}^fp_{uf}$ and $\sum_{u\notin \R_i}\textbf{a}_u^T\textbf{b}_ip_{uf}^2$, we similarly define the 3-dimensional $\textbf{S}^p$ cache, in which each element is $S^p_{tfk}=\sum_{u=1}^M a_{ut}p_{uk}p_{uf}$. With this cache, the two costly terms can be efficiently computed as:
\begin{equation}
\begin{aligned}
\sum_{u\notin \R_i} \textbf{a}_u^T\textbf{b}_i \hat{r}_{ui}^fp_{uf} &= \sum_{k\neq f}q_{ik}\sum_{t=1}^Zb_{it}\sum_{u\notin \R_i} a_{ut}p_{uk}p_{uf}\\
&= \sum_{k\neq f}q_{ik}\sum_{t=1}^Zb_{it}(S_{tfk}^p - \sum_{u\in \R_i} a_{ut}p_{uk}p_{uf})
\end{aligned}
\end{equation}
\begin{equation}
\begin{aligned}
\sum_{u\notin \R_i}\textbf{a}_u^T\textbf{b}_ip_{uf}^2 &= \sum_{u\notin \R_i}\sum_{t=1}^Z a_{ut}b_{it}p_{uf}^2\\
&= \sum_{t=1}^Zb_{it}S_{tff}^p - \sum_{u\in \R_i}\textbf{a}_u^T\textbf{b}_ip_{uf}^2
\end{aligned}
\end{equation}


Algorithm \ref{alg:ffm} summarizes the accelerated algorithm for our eALS method. 
Since each parameter update of eALS finds the optimal value for the parameter given the current status of other parameters, the training objective function is guaranteed to decrease with the training\footnote{We omit rigorous proof here since it is obvious.}. 
As such, for the stopping criteria, one can either check the objective function value, or rely on a hold-out validation data to investigate the metrics of interest. 

\begin{algorithm}
	\caption{Fast eALS algorithm for weighted MF.}
	\label{alg:ffm}
	\begin{algorithmic}[1]
		\REQUIRE{Data matrix $\textbf{R}$, number of latent factors $K$, regularizer $\lambda$, weights of observed entries $\{c_{ui}\}$, low-rank model for weights of missing entries $\textbf{A}$ and $\textbf{B}$}
		\ENSURE{MF parameters \textbf{P} and \textbf{Q}}      
		\STATE Randomly initialize $\textbf{P}$ and $\textbf{Q}$;
		\FOR{$(u,i)\in \R$}
			\STATE $\hat{r}_{ui} \leftarrow$ Eq. (\ref{eq:mf});
		\ENDFOR
		\WHILE{Stopping criteria is not met}
			\vspace{1mm}
		    \STATE // Update $\textbf{S}^q$ cache \COMMENT{$O(NK^2Z)$}
			\FORALL{$t \leftarrow 1$ \textbf{to} $Z$, $f \leftarrow 1$ \textbf{to} $K$, $k \leftarrow 1$ \textbf{to} $K$}
				\STATE $S^q_{tfk} \leftarrow \sum_{i=1}^{N}b_{it}q_{ik}q_{if}$;
			\ENDFOR
			\vspace{1mm}
			\STATE // Update $\textbf{P}$  \COMMENT{$O(MK^2Z + |\R|KZ)$}
			\FOR {{$u \leftarrow 1$ \textbf{to} $M$}}
			  \FOR {{$f \leftarrow 1$ \textbf{to} $K$}}
				\STATE \textbf{for} $i\in \R_u$ \textbf{do}  $\hat{r}_{ui}^f\leftarrow \hat{r}_{ui} - p_{uf}q_{if}$;
				\STATE $p_{uf}\leftarrow$ Eq.~(\ref{eq:p_uf})  \COMMENT{$O(|\R_u|Z+KZ)$}
				\STATE \textbf{for} $i\in \R_u$ \textbf{do} $\hat{r}_{ui}\leftarrow \hat{r}_{ui}^f + p_{uf}q_{if}$;
			  \ENDFOR
			\ENDFOR
			\vspace{1mm}
			\STATE // Update $\textbf{S}^p$ cache   \COMMENT{$O(MK^2Z)$}
			\FORALL{$t \leftarrow 1$ \textbf{to} $Z$, $f \leftarrow 1$ \textbf{to} $K$, $k \leftarrow 1$ \textbf{to} $K$}
				\STATE $S^p_{tfk} \leftarrow \sum_{u=1}^{M}a_{ut}p_{uk}p_{uf}$  
			\ENDFOR
			
			\vspace{1mm}
			\STATE // Update $\textbf{Q}$ \COMMENT{$O(NK^2Z + |\R|KZ))$}
			\FOR {{$i \leftarrow 1$ \textbf{to} $N$}}
			  \FOR {{$f \leftarrow 1$ \textbf{to} $K$}} 
				\STATE \textbf{for} $u\in \R_i$ \textbf{do}  $\hat{r}_{ui}^f\leftarrow \hat{r}_{ui} - p_{uf}q_{if}$\;
				\STATE $q_{if}\leftarrow$ Eq.~(\ref{eq:q_if})  \COMMENT{$O(|\R_i|Z+KZ)$} 
				\STATE \textbf{for} $u\in \R_i$ \textbf{do} $\hat{r}_{ui}\leftarrow \hat{r}_{ui}^f + p_{uf}q_{if}$\;
			  \ENDFOR
			\ENDFOR
		\ENDWHILE
	\end{algorithmic}
\end{algorithm}

\subsection{Discussions}
\label{ss:discussions}
In this subsection, we discuss several properties of our fast eALS algorithm, including analyzing its time complexity, the fast computation of objective function, and how to do parallel learning. 

\subsubsection{Time Complexity Analysis}\label{subsubsec:time}
The time complexities of key steps have been annotated in Algorithm \ref{alg:ffm}. Summarily, the complexity of one eALS iteration is $O((M+N)K^2Z+|\R|KZ)$, which includes the complexity of updating a row latent factor $O(KZ + |\R_u|Z)$ and the complexity of updating a column latent factor $O(KZ + |\R_i|Z)$. As we can see, even eALS models all missing data, the essential time complexity is controlled by the number of observed data $|\R|$, rather than the matrix size $M\times N$. Thus the overall time complexity is in proportion to $|\R|$ and Z, which makes eALS extremely efficient for large-scale applications. 

\begin{table}[t]
	\begin{center}
		\caption{\textbf{Time complexity of whole data-based MF methods.}}
		\vspace{-8pt}
		\label{tab:complexity}
		\begin{tabular}{ | l | l | }
			\hline
			\textbf{Method} & \textbf{Time Complexity} \\ \hline
			WALS~(Hu \etal \cite{Hu:2008})	&  $O((M+N)K^3 + |\R| K^2)$ \\ \hline
			IALS1~(Pil\'{a}szy \etal \cite{Pilaszy:2010}) & $O(K^3 + (M+N)K^2 + |\R|K)$ \\ \hline
			ii-SVD~(Volkovs \etal \cite{Volkovs:2015}) & $O((M+N)K^2 + MN\log K)$ \\ \hline
			RCD~(Devooght \etal \cite{devooght2015dynamic})	& $O((M+N)K^2 + |\R|K)$ \\ \hline
			eALS (Algorithm \ref{alg:ffm}) & $O((M+N)K^2Z + |\R|KZ)$ \\ \hline
		\end{tabular}
	\end{center}
	\scriptsize{$|\R|$ denotes the number of non-zeros in the data matrix $\textbf{R}$. $M$ and $N$ denote the number of rows and columns of data matrix $\textbf{R}$, respectively. $K$ denotes the latent dimension of MF. $Z$ denotes the rank size of the weights of missing data.}
\end{table}

There are some other MF methods that model all the missing data. Their time complexities (of one iteration) are shown in Table \ref{tab:complexity}. 
Note that besides our proposed eALS, other MF methods shown in the table only support uniform weights on missing entries. As such, there is an additional term $Z$ in our method, which denotes the rank size of the weights of missing data. 
For a fair comparison with other methods in time complexity, we assume $Z$ to be 1 and use eALS to optimize the same objective function. 
First, our model is $K$ times faster than the vector-wise ALS \cite{Hu:2008,Pan:2008}, and it has the same time complexity with RCD~\cite{devooght2015dynamic}. Moreover, it is faster than ii-SVD \cite{Volkovs:2015}, another recent solution for item recommendation with implicit feedback. It is remarkable that RCD~\cite{devooght2015dynamic} leverages the gradient descent on a randomly chosen latent vector to learn a whole-data based MF. To find out a good learning rate for faster convergence adaptively, RCD runs a line search in each gradient step. Therefore, the major advantages of eALS over RCD are the high efficiency and simplicity.

\subsubsection{Fast Computation of the Objective Function}
The value of objective function is an important indicator on the training process. A direct calculation requires evaluating every entry in the $\textbf{R}$ matrix, which takes $O(MNKZ)$ time and is very time-consuming.
To address this problem, we leverage the low-rank weighting scheme and intermediate variables cached in Algorithm \ref{alg:ffm}, devising a set of similar element-wise computations on $\textbf{R}$ for acceleration. Here, we reformulate the major cost of objective function (i.e. the loss of the missing data):
\begin{equation}
\sum_{u=1}^M\sum_{i\notin \R_u}\textbf{a}_u^T\textbf{b}_i\hat{r}_{ui}^2 = \sum_{u=1}^M\sum_{i=1}^N \textbf{a}_u^T\textbf{b}_i\hat{r}_{ui}^2 - \sum_{u=1}^M\sum_{i\in \R_u}\textbf{a}_u^T\textbf{b}_i\hat{r}_{ui}^2
\end{equation}

It is obvious that the major computation comes from the first term, due to the iterations over all rows and columns.  
Thus we accelerate it with the transformation in Eq.~(\ref{eq:trans}):
\begin{equation} \small
\label{eq:trans}
\begin{aligned}
\sum_{u=1}^M\sum_{i=1}^N\textbf{a}_u^T\textbf{b}_i\hat{r}_{ui}^2 
&= \sum_{u=1}^M\sum_{i=1}^N\sum_{t=1}^Z a_{ut}b_{it} \sum_{k=1}^Kp_{uk}q_{ik}\sum_{f=1}^K q_{if}p_{uf} \\
&= \sum_{u=1}^M\sum_{k=1}^Kp_{uk}\sum_{f=1}^Kp_{uf}\sum_{t=1}^Za_{ut}\sum_{i=1}^N(b_{it}q_{ik}q_{if}) \\
&= \sum_{u=1}^M\sum_{k=1}^Kp_{uk}\sum_{f=1}^Kp_{uf}\sum_{t=1}^Za_{ut}S_{tfk}^q.
\end{aligned}
\end{equation}
As can be seen, with the help of the $\textbf{S}^q$ cache, we can reduce the time complexity to $O(MK^2Z)$, which is several orders of magnitude smaller than the direct computation of $O(MNKZ)$. 

\subsubsection{Parallel Learning}
The key operations of eALS are easily parallelizable. First, the computations of the two caches $\textbf{S}^q$ and  $\textbf{S}^p$ are based on the standard matrix multiplication operations (line 8 and 20), which are straightforward to be parallelized. Second, in updating the latent vectors $\textbf{P}$ (line 10-17), the cache $\textbf{S}^q$ is temporarily fixed, and the shared parameters $\hat{r}_{ui}$ are independent with each other. 
Therefore, it is practicable to update rows in parallel due to the nice independent property. Specifically, eALS can leverage multiple workers to update the model parameters for disjoint sets of rows concurrently. Similarly, this parallel strategy can also be applied in updating latent vectors $\textbf{Q}$ (line 22-29). 

It is notable that since the operations in SGD are strictly ordered and they are hard to be separated, controlling the possible losses is significant and difficult to leverage sophisticated strategies in parallel~\cite{gemulla2011large}. Meanwhile, the SGD loss always constrains the paralleling magnitude. Thus the ease of parallelization is an important advantage of our proposed eALS over the commonly used SGD learner. With coordinate descent, by parallelizing the key operations, our proposed eALS is embarrassingly parallel without any approximation loss.

\section{Experiments}\label{sec:experiments}
In this section, we perform experiments to verify the correctness, efficiency, and effectiveness of our fast eALS algorithm. All experiments are conducted on two real-world rating datasets, which are commonly used in recommendation systems. We first introduce the experimental settings, followed by the verification of correctness and efficiency, and the effectiveness of eALS in recommendation by modeling missing data with non-uniform weights. Note that part A, D and Table IV of this section have been presented in the preliminary version \cite{he2016fast}, and other parts are new. 

\subsection{Experimental Settings}\label{ss:settings}
\subsubsection{Datasets}
Two publicly accessible rating datasets are selected to evaluate the methods: Yelp\footnote{We used the Yelp Challenge dataset downloaded on October 2015 that contained 1.6 million reviews.} and Amazon Movies\footnote{\url{http://snap.stanford.edu/data/web-Amazon-links.html}}. 
The Yelp dataset is about users' ratings on businesses (most of which are restaurants) and the Amazon dataset is about users' ratings on movies, where each rating is in the range of 1 to 5.
We construct the data matrix $\textbf{R} \in \mathbb{R}^{M\times N}$ by defining each row as a user, each column as an item, and each entry as the user's rating score on the item; if a user did not rate an item before, the corresponding entry in $\textbf{R}$ will be defined as missing data. 
Table~\ref{tab:dataset} summarizes the statistics of our experimented datasets\footnote{The experimented datasets can be downloaded from: \url{https://github.com/hexiangnan/sigir16-eals/tree/master/data}}. We can see that both datasets are extremely sparse with the sparsity ratio over $99.8\%$. This provides empirical evidence on the necessity of using low-rank weights rather than storing the whole weight matrix as it is. Take the Amazon dataset as an example, storing the whole dataset matrix takes the space of $35 GB$ ($75,389 * 117,176 * 4B\approx35 GB$), which is very space-consuming.  

\begin{table}
	\begin{center}
		\caption{\textbf{Statistics of the evaluation datasets.}}
		\vspace{-8pt}
		\small
		\label{tab:dataset}
		\begin{tabular}{ | l | c | c | c | c | }
			\hline
			\textbf{Dataset} & \textbf{Review\#} & \textbf{Item\#} & \textbf{User\#}  & \textbf{Sparsity} \\ \hline
			Yelp	& 731,671	& 25,815 & 25,677  & 99.89\% \\ \hline
			Amazon	& 5,020,705	& 75,389 & 117,176  & 99.94 \% \\ \hline
		\end{tabular}
		\vspace{-20pt}
	\end{center}
\end{table}

\subsubsection{Evaluation Protocols}
We split the data into training set and testing set by using the leave-one-out protocol, a widely used method in recommendation papers \cite{Rendle:2009:BPR,APR}. Specifically, the last rating of each user is hold out for testing, and the remaining data are used for training. 

Besides validating the model training with the loss value, we also evaluate the performance of \textit{Hit Ratio} (HR) and \textit{Normalized Discounted Cumulative Gain}~(NDCG), which judge the ranking quality of top-N recommendation; here we set the N to 100 without special mention. More specifically, for a user, we rank all unrated items by its prediction scores, and take out the top-100 items as the \textit{recommended list}. If the testing item appears in the recommended list, it is treated as a hit, and the HR is set as 1. We can see that HR does not account for the position of a hit --- as long as the testing item appears in the recommended list, it is treated as a success. NDCG addresses this deficiency by assigning higher rewords to hits at top positions and scoring successively lower-position hits with marginal fractional utility. More details about the two metrics can be found in \cite{he2016fast}.

Since the evaluation on each user produces a ranking list, we calculate HR and NDCG for each user and report the average score of all users. Clearly, a higher score denotes a better performance, and both measures are in the range of 0 to 1. 

\subsection{Correctness Verification}
Here we verify the correctness of our proposed eALS algorithm. Since SVD is known to optimize the unweighted squared loss and can find the global minimum, we setup eALS to optimize the same objective function as SVD to verify its correctness. Specifically, we set $c_{ui}$ for all observed entries as 1, the regularization parameter $\lambda$ to 0, and $\textbf{A}$ and $\textbf{B}$ to a vector in which all entries are 1; the number of latent factors $K$ is set to 64 for both methods. For SVD, we use the Python toolkit sparsesvd\footnote{\url{https://pypi.python.org/pypi/sparsesvd/}}, which solves the SVD problem with the LBD method. For eALS, we run it for 100 iterations.

\begin{table}[t]
	\begin{center}
		\caption{\textbf{Comparison between SVD and our eALS algorithm in optimizing the same objective function.}}
		\vspace{-8pt}
		\label{tab:svd}
		\begin{tabular}{ | l | l | c | c | c |}
			\hline
			\textbf{Dataset} & \textbf{Approach}  & \textbf{HR} & \textbf{NDCG}  & \textbf{Training Loss}\\\hline
			Yelp & SVD		& 0.1816 & 0.0439 & $5.9222*10^5$
			\\\hline
			& eALS 	& 0.1814 & 0.0438 & $5.9223*10^5$ \\\hline
			Amazon & SVD  & 0.2945 & 0.0690 & $3.0172*10^6$\\\hline
			& eALS	& 0.2945 & 0.0690 & $3.0172*10^6$\\\hline
		\end{tabular}
	\end{center}
\end{table}

\begin{figure}[t]
	\centering
	\includegraphics[width=\linewidth]{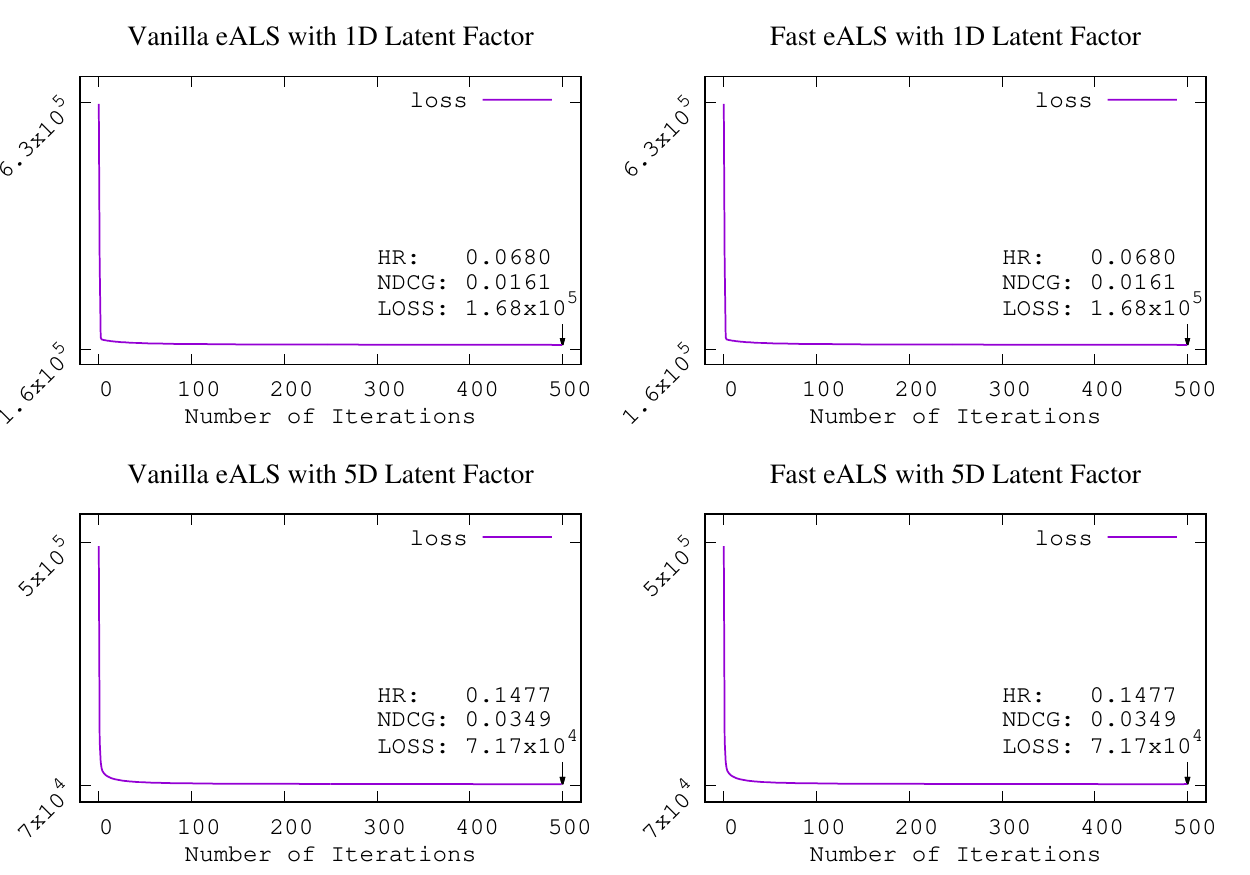} 
	\caption{The comparison between the vanilla eALS (Algorithm \ref{alg:gefm}) and fast eALS (Algorithm \ref{alg:ffm}) in terms of training loss and testing scores. The left two subfigures show vanilla eALS with K equals to 1 and 5, respectively, and the right two subfigures show fast eALS with the same setting.}
	\label{fig:effect}
\end{figure}

Table \ref{tab:svd} shows the loss achieved by the two methods on the training set and their HR and NDCG scores on the testing set. We can see that eALS achieves almost the identical performance as SVD. On Yelp, the training loss of eALS is slightly higher than that of SVD, which is caused by the insufficient training of eALS, since eALS iteratively updates model parameters while SVD finds the global optima with a closed form solution; and the t-tests show that the two methods are in the same significance level. On Amazon, eALS sufficiently converges in 100 iterations, and both training loss and testing scores show that eALS achieves the same performance as SVD. 
To verify that eALS finds the exactly same solution as SVD, we further employ the point-wise measure \textit{ mean absolute error} (MAE) to evaluate the difference of the two methods' prediction on observed entries: on Yelp, the MAE is $3.1*10^{-4}$; and on Amazon, the MAE is $9.7*10^{-6}$. Such tiny error rate is acceptable, considering that eALS and SVD are implemented with different programming languages with different float precision settings.  
Overall, these results verify the correctness of our fast eALS algorithm.

\begin{figure*}[t]
	\centering
	\subfloat[Training time per iteration in Yelp]{\includegraphics[width=0.48\linewidth]{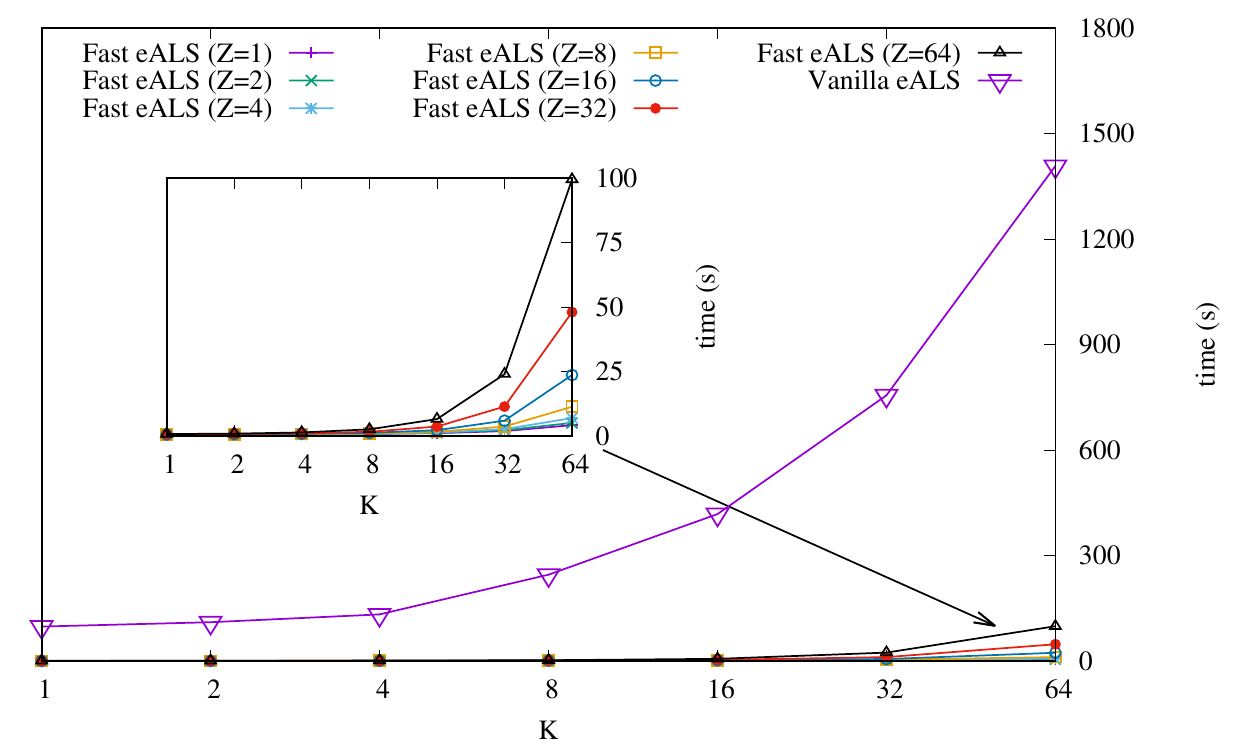}\label{fig:eff_yelp}}
	\subfloat[Training time per iteration in Amazon]{\includegraphics[width=0.48\linewidth]{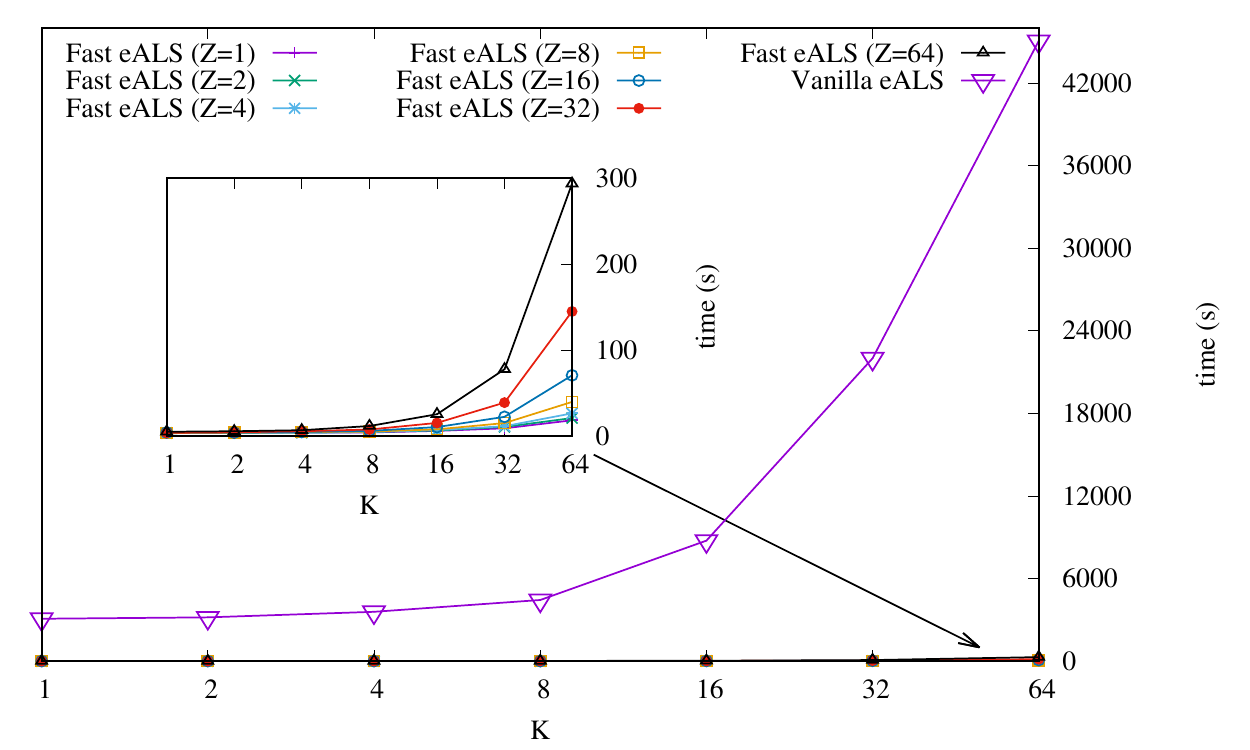}\label{fig:eff_amazon}}
	\caption{Comparison on the training time of the vanilla eALS algorithm and fast eALS (with different settings of $Z$).}
	\label{fig:efficiency}
\end{figure*}

Furthermore, we empirically test whether our fast eALS method (Algorithm~\ref{alg:ffm}) speeds up the vanilla eALS (Algorithm~\ref{alg:gefm}) without any sacrifice on the accuracy. 
To avoid the possible randomness that affects the results, we apply the same initialization on their model parameters . Figure \ref{fig:effect} shows the training process of the two algorithms with the number of latent factors $K$ setting to 1 and 5 on Yelp. We can see that the fast eALS algorithm obtains exactly the same result as the vanilla eALS, including the training loss of each epoch. This is as expected, since we derive the fast eALS based on rigorous mathematical operations without any approximation. This further justifies the correctness of the fast eALS algorithm, which is actually non-trivial to implement since it has several caches for speed-up purpose that need to be carefully updated. Interested readers can check out our implementation at: \url{https://github.com/duxy-me/ext-als}.

\subsection{Efficiency Study}
We investigate the actual speedup brought by our design of the fast eALS algorithm. All experiments in this subsection are run on the same machine (Intel Xeon 2.67GHz CPU and 24GB RAM) for fair comparison on the efficiency. Figure~\ref{fig:efficiency} shows the training time of the vanilla eALS and our fast eALS with different settings of $Z$ and $K$. In the figure, the x-axis denotes the setting of $K$, the y-axis denotes the training time per iteration, and different lines indicate different settings of $Z$. We have the following key observations:
\begin{itemize}
\item The fast eALS is several magnitudes faster than the vanilla eALS algorithm. For example, in Yelp, the vanilla eALS takes 98 seconds to train a small model of $K=1$, while the fast eALS takes only 0.8 seconds to train the same model even with a large $Z$ of 64. The speedup is more significant for the larger Amazon data, where the vanilla eALS takes over 42,000 seconds (i.e., half day) to train a model of $K=64$, while the fast eALS takes 
only 300 seconds to train the same model with a large $Z$ of 64. This acceleration is over 100 times, which is highly valuable in practice and is difficult to achieve with simply engineering efforts. Intuitively, one needs to have over 100 machines and implements an effective distributed system with a negligible network cost and linear scale-up on the number of machines, which is very difficult to achieve in practice~\cite{Rendle:2016}.
\item The running time of fast eALS exhibits a linear relationship with respect to $Z$, which can be seen clearly from the inside box of the figure. For example in Yelp, for $K=64$, the y-axis of $Z=64$ is twice of that of $Z=32$, which is the same for the Amazon dataset. Moreover, the running time exhibits a quadratic relationship with respect to $K$. These results are as expected, verifying the analytical time complexity of the fast eALS algorithm --- $O((M+N)K^2Z + |\R|KZ)$.
\end{itemize}

\noindent Furthermore, we compare the efficiency of the fast eALS algorithm with two whole data-based MF methods --- the 
\textit{Randomized block Coordinate Descent} (RCD) \cite{devooght2015dynamic} method and the WALS method~\cite{Hu:2008}. Since both methods only support uniform weighting on missing data, we set the $Z$ of eALS to 1 for a fair comparison. Table \ref{tab:efficency} shows the average training time per iteration of the three methods. 

\begin{table}[h]
	\begin{center}
		\caption{\textbf{Training time per iteration of fast eALS, RCD and WALS.}}
		\small
		\vspace{-8pt}
		\label{tab:efficency}
		\begin{tabular}{ | c | c | c | c | c | c | c |}
			\hline
			& \multicolumn{3}{c}{\textbf{Yelp}} & \multicolumn{3}{|c|}{\textbf{Amazon}} \\ \hline 
			\textbf{K} & \textbf{eALS} & \textbf{RCD} & \textbf{WALS} & \textbf{eALS} & \textbf{RCD} & \textbf{WALS} \\ \hline
			\textbf{32}	& 1s & 1s & 10s & 9s & 10s & 74s \\ \hline
			\textbf{64}	& 4s & 3s & 46s & 23s & 17s & 4.8m \\ \hline
			\textbf{128}	& 13s & 10s & 221s & 72s & 42s & 21m \\ \hline
			\textbf{256}	& 1m& 0.9m& 23m & 4m & 2.8m & 2h \\ \hline
			\textbf{512}	& 2m& 2m & 2.5h & 12m & 9m & 11.6h \\ \hline
			\textbf{1024}	& 7m& 11m & 25.4h & 54m & 48m & 74h  \\ \hline
		\end{tabular}
	\end{center}
	\scriptsize{\quad$s$, $m$, and $h$ denote seconds, minutes and hours, respectively.}
	\vspace{-5pt}
\end{table}

Analytically, WALS has the time complexity of $O((M+N)K^3 + |\R|K^2)$, while eALS and RCD have the same time complexity which is $K$ times smaller than that of WALS. As can be seen from the table, with the increase of $K$, WALS takes much longer time than eALS and RCD. Specifically, when $K$ is 512, WALS requires 11.6 hours for one iteration on Amazon, while eALS only takes 12 minutes. 
Although eALS does not empirically shown to be $K$ times faster than ALS due to the more efficient matrix inversion implementation (we used the fastest known algorithm \cite{Coppersmith:1990} with time complexity around $O(K^{2.376})$), the speed-up is already very significant.
Moreover, as RCD and eALS have the same analytical time complexity, their actual running time are in the same magnitude; the minor differences can be caused by some implementation details, such as the data structures used. 

\begin{figure*}[t]
	\centering
	\subfloat[eALS vs. $c_0\ (\alpha=0)$]{\includegraphics[width=0.25\textwidth]{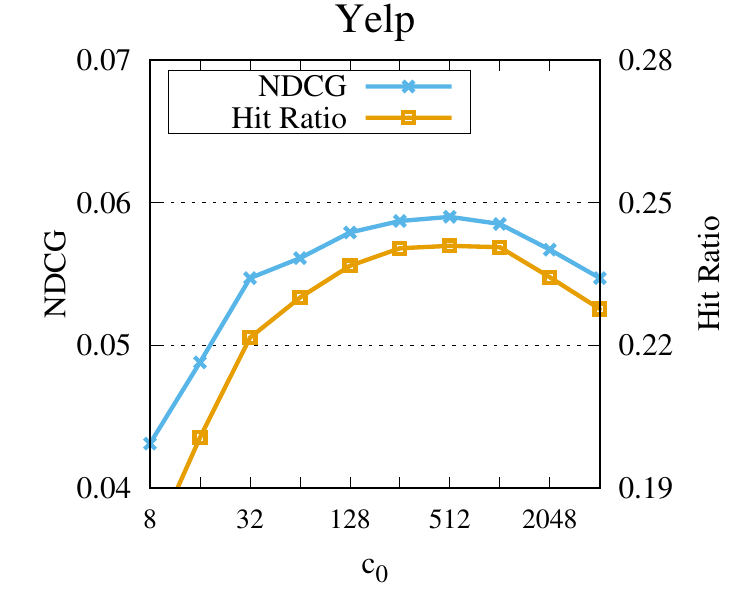}\label{fig:yelp_w0}}
	\subfloat[eALS vs. $\alpha\ (c_0=512)$]{\includegraphics[width=0.25\textwidth]{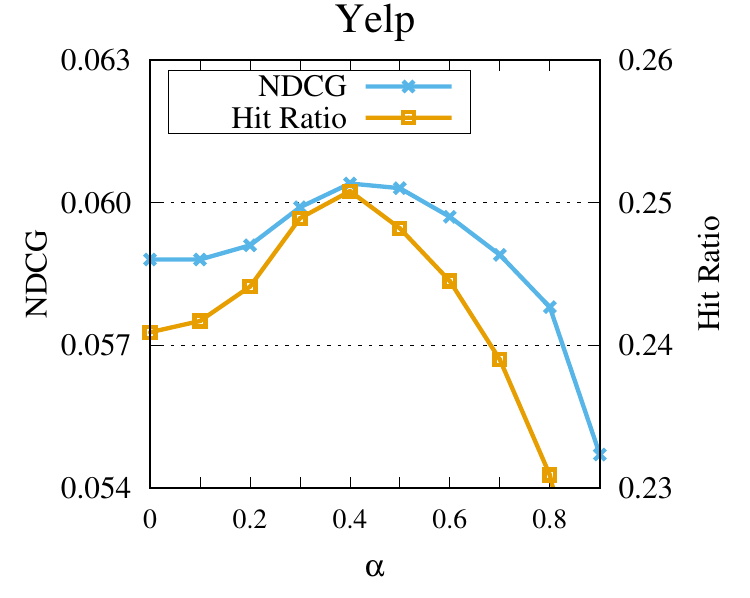}\label{fig:yelp_alpha}}
	\subfloat[eALS vs. $c_0\ (\alpha=0)$]{\includegraphics[width=0.25\textwidth]{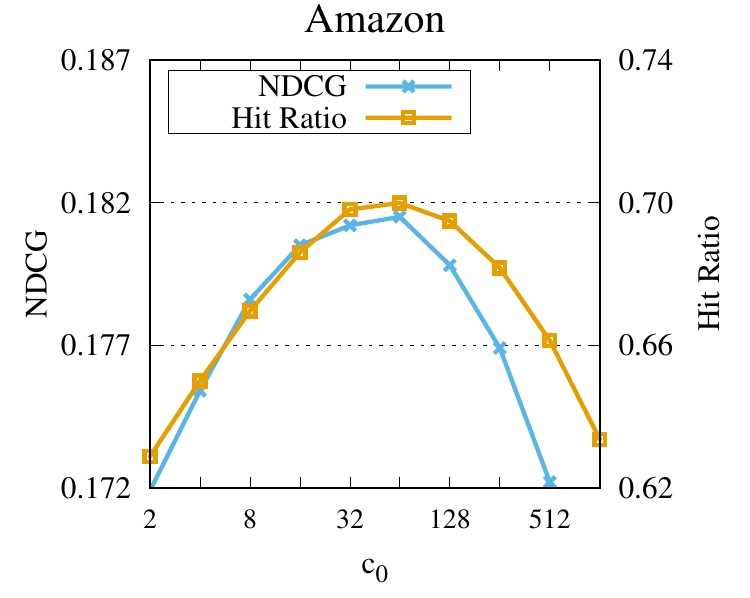}\label{fig:amazon_w0}}
	\subfloat[eALS vs. $\alpha\ (c_0=64)$]{\includegraphics[width=0.25\textwidth]{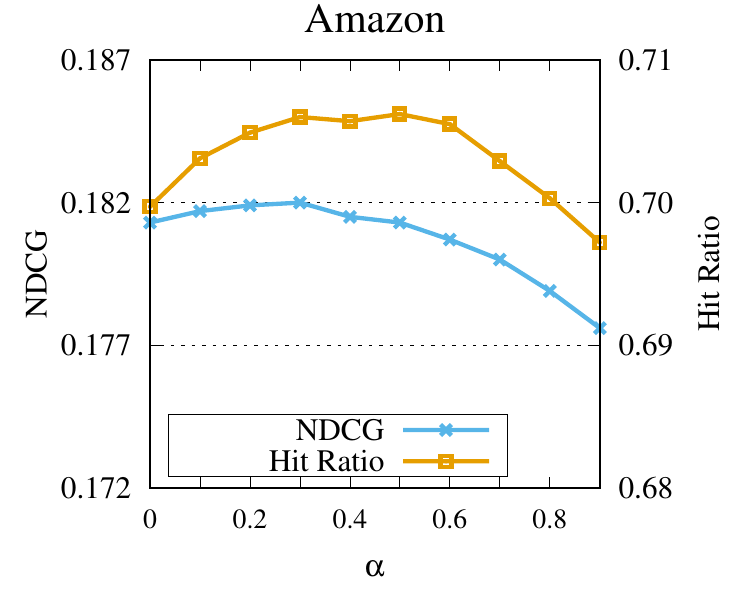}\label{fig:amazon_alpha}}
	\caption{Performance of eALS with respect to $c_0$ and $\alpha$ on Yelp and Amazon. Each subfigure shows the scores of varying one parameter with the other fixed.}
	\label{fig:weight}
\end{figure*}

\subsection{Effectiveness in Item Recommendation} \label{ss:recom_results}

In this subsection, we explore the effectiveness eALS in the real-world task of item recommendation. As mentioned in Section~\ref{ss:settings}, this is a personalized ranking task, and we employ the leave-out-one protocol to evaluate the performance with NDCG and HR. We aim to answer the following research questions:
\begin{itemize}
	\item[\textbf{RQ1}] Is the non-uniform weighting strategy on missing data effective to offer better performance?
	\item[\textbf{RQ2}] How does eALS perform as compared with existing whole data-based MF methods for recommendation?
\end{itemize}
Next, we describe experimental results to answer the two questions. Note that our findings are consistent across the number of latent factors $K$, thus we show the results of $K=128$ only, a relatively large number that retains good model capability. 

\textbf{RQ1: Non-Uniform Weights on Missing Data}. Existing MF methods for recommendation assign a uniform weight on missing data for the ease of efficient optimization. This implies that all unrated items for a user have an equal probability to be negative, which may not be true. Since the visual interfaces of many Web systems tend to showcase popular items, when all other factors are equal, popular items are more likely to be known by users in general~\cite{He:SIGIR14}. As such, it is reasonable to think that a miss on a popular item is more probable to be truly irrelevant (as opposed to unknown) to the user. To account for this effect, we design the weights for missing entries based on item popularity:
\begin{equation}
w_{ui} = c_0 \frac{f_i^\alpha}{\sum_{j=1}^N f_j^\alpha},
\end{equation}
where $f_i$ denotes the frequency of item $i$, in the training set: $|\R_i|/\sum_{j=1}^N{|\R_j|}$.
The weight for each observed entry (i.e., $c_{ui}$ in Eq.~(\ref{eq:partitioned})) is set as 1, and $c_0$ is a hyper-parameter to determine the overall weight of missing data. 
The exponent $\alpha$ controls the significance level of popular items over unpopular ones --- when $\alpha > 1$ the weights of popular items are promoted to strengthen the difference against unpopular ones; while setting $\alpha$ within the lower range of $(0,1)$ suppresses the weight of popular items and has a smoothing effect. This weighting strategy basically assumes that the missing entries of popular items carry more negative signal. 
It is obvious that the rank size of such a weight matrix (on missing entries only) is 1, and with truncated SVD, we can get its low-rank representation as:
\begin{equation}\label{eq:neg_scheme}
\begin{cases}
\textbf{a}_{u} = [c_0] \\
\textbf{b}_i = [\frac{f_i^\alpha}{\sum_{j=1}^N f_j^\alpha}].
\end{cases}
\end{equation}
Our fast eALS implementation is initialized with this setting on the weights of missing entries. 
It is worth noting that other non-uniform weighting strategies can also be applied here, such as the user-oriented scheme proposed in \cite{li2018cf}, or we can combine user-oriented and item-oriented schemes in Eq. (\ref{eq:neg_scheme}). Since the aim of this experiment is to show the effectiveness of customizing weights in eALS, rather than demonstrating state-of-the-art recommendation performance, we leave this further exploration on the weighting scheme as future work. 
Note that this initialization leads to a recommendation method same as our preliminary work \cite{he2016fast}. As such, the results presented below are also the same.  

\noindent\textbf{Results.} Figure \ref{fig:weight} shows the performance of eALS with respect to $c_0$ and $\alpha$. Let us first focus on Figure \ref{fig:yelp_w0} and \ref{fig:amazon_w0}, where the weights of missing data follow a uniform distribution (controlled by $\alpha=0$); we vary $c_0$ to study how does the overall weight of missing data affect the performance. The optimal $c_0$ on Yelp (Figure~\ref{fig:yelp_w0}) is around 512, and that on Amazon (Figure~\ref{fig:amazon_w0}) is around 64. Correspondingly, the weights of each zero entry are $0.02$ and $0.0001$ respectively ($w_0=c_0/N$). However, both datasets exhibit similar patterns: when $c_0$ is smaller than the optimal value, the performance drops significantly. In other words, when the weights of zero entry are close to 0, the performance degrades. That reflects the importance of weighting the missing data. Moreover, too large $c_0$ also leads to bad performance. That is why the traditional SVD technique~\cite{Cremonesi:2010}, which assigns the same weight to all the entries, is suboptimal here.

\begin{figure*}[t]
	\centering
	\subfloat[Iterations vs. HR]{\includegraphics[width=0.25\textwidth]{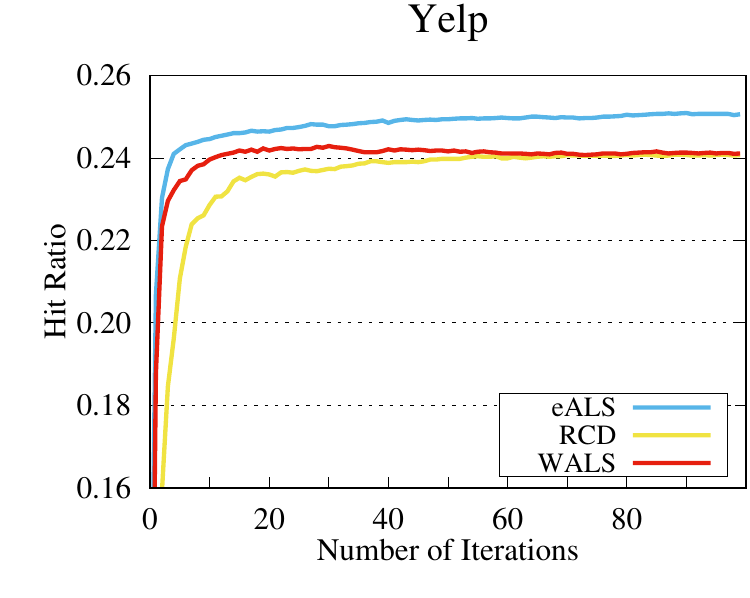}\label{fig:yelp_mf_hr}}
	\subfloat[Iterations vs. NDCG]{\includegraphics[width=0.25\textwidth]{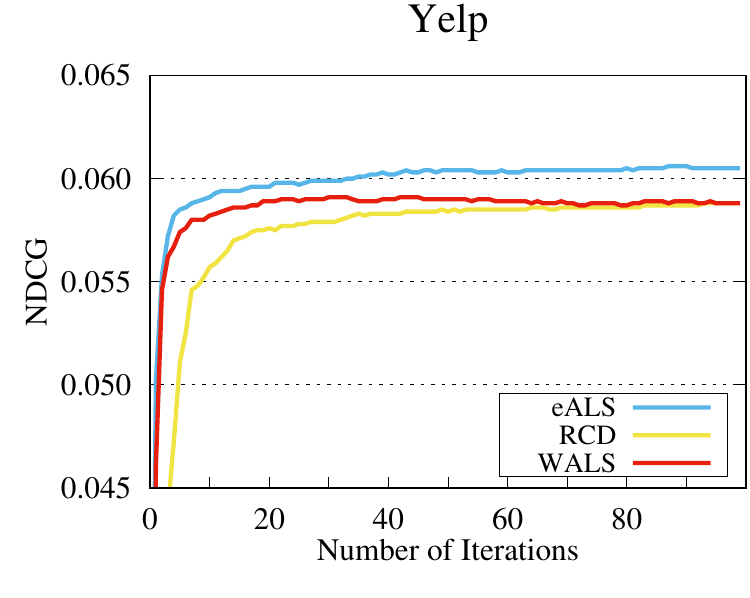}\label{fig:yelp_mf_ndcg}}
	\subfloat[Iterations vs. HR]{\includegraphics[width=0.25\textwidth]{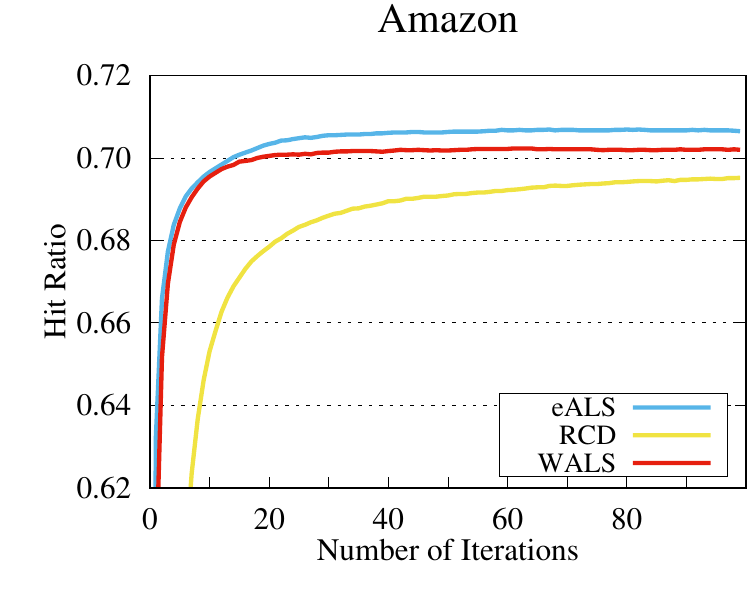}\label{fig:amazon_mf_hr}}
	\subfloat[Iterations vs. NDCG]{\includegraphics[width=0.25\textwidth]{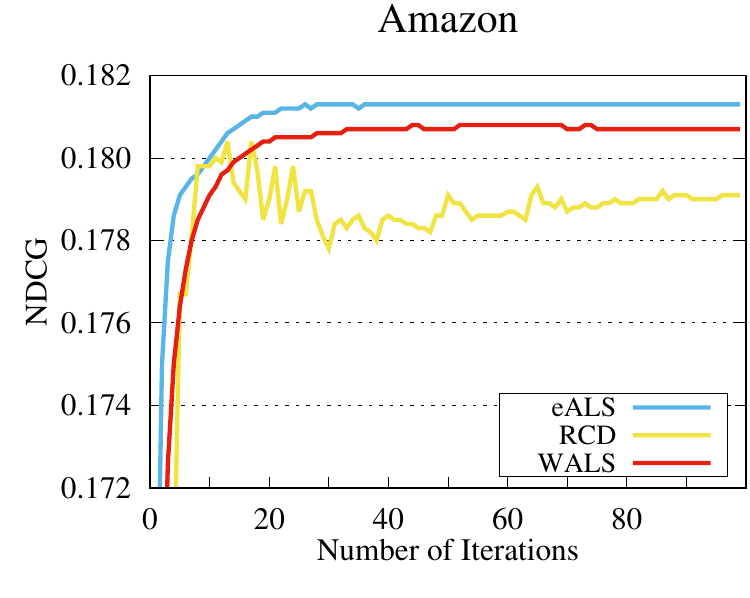}\label{fig:amazon_mf_ndcg}}
	\caption{Recommendation accuracy in each training iteration of three whole-data based MF methods, eALS, RCD and WALS ($K=128$).}
	\label{fig:mf_convergence}
\end{figure*}

Then, we vary $\alpha$ with the optimal $c_0$ (in the case of $\alpha=0$) to check the performance change.  
As demonstrated in Figure~\ref{fig:yelp_alpha} and \ref{fig:amazon_alpha}, the optimal $\alpha$ is around 0.4 on both datasets. Below 0.4, with the increase of $\alpha$, the performance of eALS is gradually improved. But when $\alpha$ increases above 0.5, the performance of eALS becomes worse.  That reveals that weighting missing data according item popularity is important for recommendation. We further verify the improvement with the one-sample paired $t$-test. The results ($p$-value $< 10^{-4}$) for both metrics on the two datasets indicates the effectiveness of our method.

In the following experiments, we fix $c_0$ and $\alpha$ according to the best performance evaluated by HR, i.e., $c_0=512, \alpha=0.4$ for Yelp and $c_0=64, \alpha=0.5$ for Amazon.

\textbf{RQ2: Performance Comparison}.
We compare eALS with two whole data-based MF methods which are originally designed for the item recommendation task:

- \textbf{WALS}~\cite{Hu:2008}. This is the weighted ALS method that optimizes the whole-data based MF. It assigns the same weight $w_0$ to all missing data. We tuned the $w_0$ carefully and reported the best performance. 

- \textbf{RCD}~\cite{devooght2015dynamic}. This is the state-of-the-art implicit MF method that optimizes the same objective function as ALS but with a faster coordinate descent learner. We similarly tuned the $w_0$; for the line search related parameters, we use the suggested values in the authors' implementation\footnote{\scriptsize{\url{https://github.com/rdevooght/MF-with-prior-and-updates}}}.

The recommendation accuracy of each training iteration is shown in Figure~\ref{fig:mf_convergence}.
All improvements are statistically significant as evidenced by the one-sample paired $t$-test ($p<0.01$). First of all, the performance of eALS is best upon convergence. We believe that the improvements are mainly from the weighting strategy on the missing data.  Our proposed eALS assigns adaptive weights to the missing data while both WALS and RCD apply uniform weights. 

Second, RCD converges slower than eALS and WALS. This is caused by the difference between global optimization and local optimization. In each iteration, RCD updates to a suboptimal point, which may be a wrong direction to the global optimal point. The effect of local optimization is also verified in Figure~\ref{fig:amazon_mf_ndcg}. RCD attains high NDCG in a short time, while its low HR and the later oscillation demonstrate that the high NDCG of RCD is unstable. Another reason for this situation may come from the RCD's adaptive strategy. In the early iterations, the optimizer tends to make rapid learning by using a large learning rate, that may lead to the unexpected (i.e. suboptimal) results. Nevertheless, WALS outperforms RCD in most situations,  that demonstrates ALS is better than gradient descent learner in these tasks.

\section{Related work}
Matrix Factorization is a representative method that represent a data matrix as two low-dimension matrices.
The decomposition process can distill co-occurrence patterns in data~\cite{DBLP:tnn/MaXLTYG18}. Moreover, the reconstructed low-rank model can be used to recover missing information, such as its application in recommendation that predicts users' ratings on unknown items~\cite{TNNLS16Recom,DCF:2016}. 

However, in many real-world applications, the data matrices are can be highly sparse. 
For example, in recommendation, handling missing data is particularly important for learning from implicit data, since they provide valuable negative signal. Along this line, we can categorize previous works into two types: \textit{sample-based learning}  and \textit{whole-data based learning}: 

- The first type samples negative instances from missing data~\cite{Rendle:2009:BPR,he2017neural,ConvNCF,zhang2017joint}. For example, the BPR method proposed by Rendle et al.~\cite{Rendle:2009:BPR} randomly samples negative instances from missing entries, maximizing the margin between the model prediction of observed entries and that of sampled negatives. 
Recently, He et al.~\cite{APR} develops adversarial training methods for BPR to increase the robustness of the learned model. 
By negative sampling, the number of negative instances is greatly reduced, therefore the overall time complexity is controllable~\cite{he2017neural}. 
However, the downside is that they usually have a slower convergence rate and the performance is highly dependent of the design of the sampler~\cite{AllVec,Ding:2018,Rendle:2014:IPL}.

- The second type treats all missing entries as negative instances ~\cite{Hu:2008,iCD,yuan2018fbgd,DBLP:ijcai/DingY0QLCJY18}. For example, the WALS method proposed by Hu et al.~\cite{Hu:2008} models all missing entries as negative instances with a label of 0, assigning them with a lower weight in point-wise regression learning.
Recently, Ding et al.~\cite{DBLP:ijcai/DingY0QLCJY18} develops a pairwise learning framework to model the margin between observed entries (based on view histories) and all missing entries.   
These methods model negative instances with a higher coverage, but the downside is that the learning algorithm could be much slower. 

To pursue model effectiveness, we focus on whole-data based learning in this work, aiming to develop an efficient solution to address the inefficiency issue. For this line of research, several previous efforts have been made, such as \cite{iCD,devooght2015dynamic,Hu:2008,Pilaszy:2010,iCD,Volkovs:2015}. 
We find that these methods have a common limitation --- weighting missing entries with a same weight. 
This design is mainly for efficiency concern, since fast learning algorithms can be obtained with this constraint. However, it decreases the modeling flexibility and may result in suboptimal performance. 
The works that are closest to ours are~\cite{Pan:2008,yuan2018fbgd,li2018cf}, which consider applying non-uniform weights on missing entries. 
However, these methods only supports simple weighting scheme, either row-based or column-based, and cannot be extended to other more complex schemes. This work addresses the research gap by developing efficient learning algorithms for any weighting scheme on missing data. Lastly, it is worth noting that the algorithm proposed in the recent work \cite{li2018cf}
is a special case of our fast eALS method, since it can be exactly recovered by setting the weights on missing entries to be user-oriented.

\section{Conclusion}
In this paper, we studied the problem of learning MF with non-uniform weights on missing data. Targeting at the $L_2$ square loss, we first proposed to apply ALS optimization at each element level, namely, eALS. To address the efficiency challenge in solving the weighted MF problem, we then proposed a low-rank weighting strategy on missing data, which not only saves the space in storing weights but also allows us to further speedup the eALS method. To this end, we developed a fast eALS algorithm by a clever use of memoization caches, for which the time complexity is determined by the number of observed entries only rather than the whole data matrix. 
We conducted extensive experiments on two public rating datasets, verifying the correctness, efficiency, and effectiveness of our proposed fast eALS method. 

We believe that optimizing MF with missing data is a fundamental problem in learning on sparse matrices. While most existing works assign a uniform weight on missing data, this work opens the door for designing complex weighting schemes for missing data. This will benefit a wide variety of tasks that can be solved with MF. In future, we plan to extend eALS to MF with side information, such as spatial contexts~\cite{yang2017bridging}, user reviews~\cite{He:2015}, visual content~\cite{Wang:2017:YIR}, and knowledge graphs~\cite{ai2018learning}. Moreover, we will consider applying non-uniform weights for missing data on the more generic embedding models, such as collective factorization~\cite{He:WWW2014} and neural factorization machines~\cite{NFM}. Lastly, we are interested in applying our method on other tasks, such as the knowledge graph completion and word representation learning.

\appendices
\ifCLASSOPTIONcompsoc
  \section*{Acknowledgments}
\else
  \section*{Acknowledgment}
\fi

The authors thank the anonymous reviewers for their reviewing efforts. This research is supported by the National Natural Science Foundation of China (Grant No. 61772275, 61732007, 61321491, 61202320, 61501063), the Outstanding Youth Science Foundation (No. 61722204), the Scientific Research Foundation of Science and Technology Department of Sichuan Province(Grant No.2016JY0240), and the Collaborative Innovation Center of Novel Software Technology and Industrialization. This research is also part of NExT++, supported by the National Research Foundation, Prime Ministers Office, Singapore under its IRC@Singapore Funding Initiative.
This work is a significant extension
of \cite{he2016fast}, which appeared in the \textit{Proceedings of SIGIR 2016}.

\ifCLASSOPTIONcaptionsoff
  \newpage
\fi

\bibliographystyle{IEEEtran}
\bibliography{bib}
\begin{IEEEbiography}[{\includegraphics[width=1in,height=1.25in,clip,keepaspectratio]{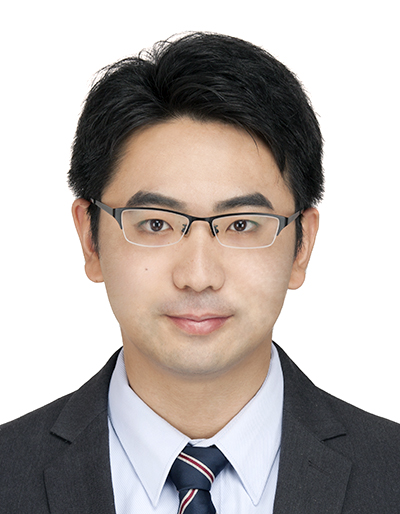}}]{Xiangnan He} is currently a professor with the School of Information Science and Technology, University of Science and Technology of China (USTC). He received his Ph.D. in Computer Science from National University of Singapore (NUS) in 2016, and did postdoctoral research in NUS until 2018. His research interests span information retrieval, data mining, and multi-media analytics. He has over 50 publications appeared in several top conferences such as SIGIR, WWW, and MM, and journals including TKDE and TOIS. His work on recommender systems has received the Best Paper Award Honourable Mention in WWW 2018 and ACM SIGIR 2016. Moreover, he has served as the PC member for several top conferences including SIGIR, WWW, MM, KDD etc., and the regular reviewer for journals including TKDE, TOIS, TNNLS etc.
\end{IEEEbiography}\vspace{-20pt}

\begin{IEEEbiography}[{\includegraphics[width=1in,height=1.25in,clip,keepaspectratio]{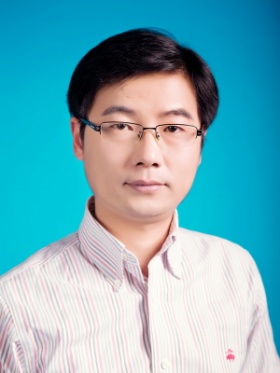}}]{Jiuhui Tang} is currently a Professor in School of Computer Science and Engineering, Nanjing University of Science and Technology, China. He received the B.Eng. and Ph.D. degrees from the University of Science and Technology of China, Hefei, China, in 2003 and 2008, respectively. From 2008 to 2010, he worked as a research fellow in School of Computing, National University of Singapore. His current research interests include multimedia content analysis and retrieval, social media mining and machine learning. He has authored over 100 papers in top-tier journals and conferences. Dr. Tang is a recipient of the inaugural ACM China Rising Star Award, the Best Paper Awards in ACM MM 2007, PCM 2011 and ICIMCS 2011, the Best Paper Runner-up in ACM MM 2015, and the Best Student Paper Awards in MMM 2016 and ICIMCS 2017. 
\end{IEEEbiography}\vspace{-20pt}

\begin{IEEEbiography}[{\includegraphics[width=1in,height=1.25in,clip,keepaspectratio]{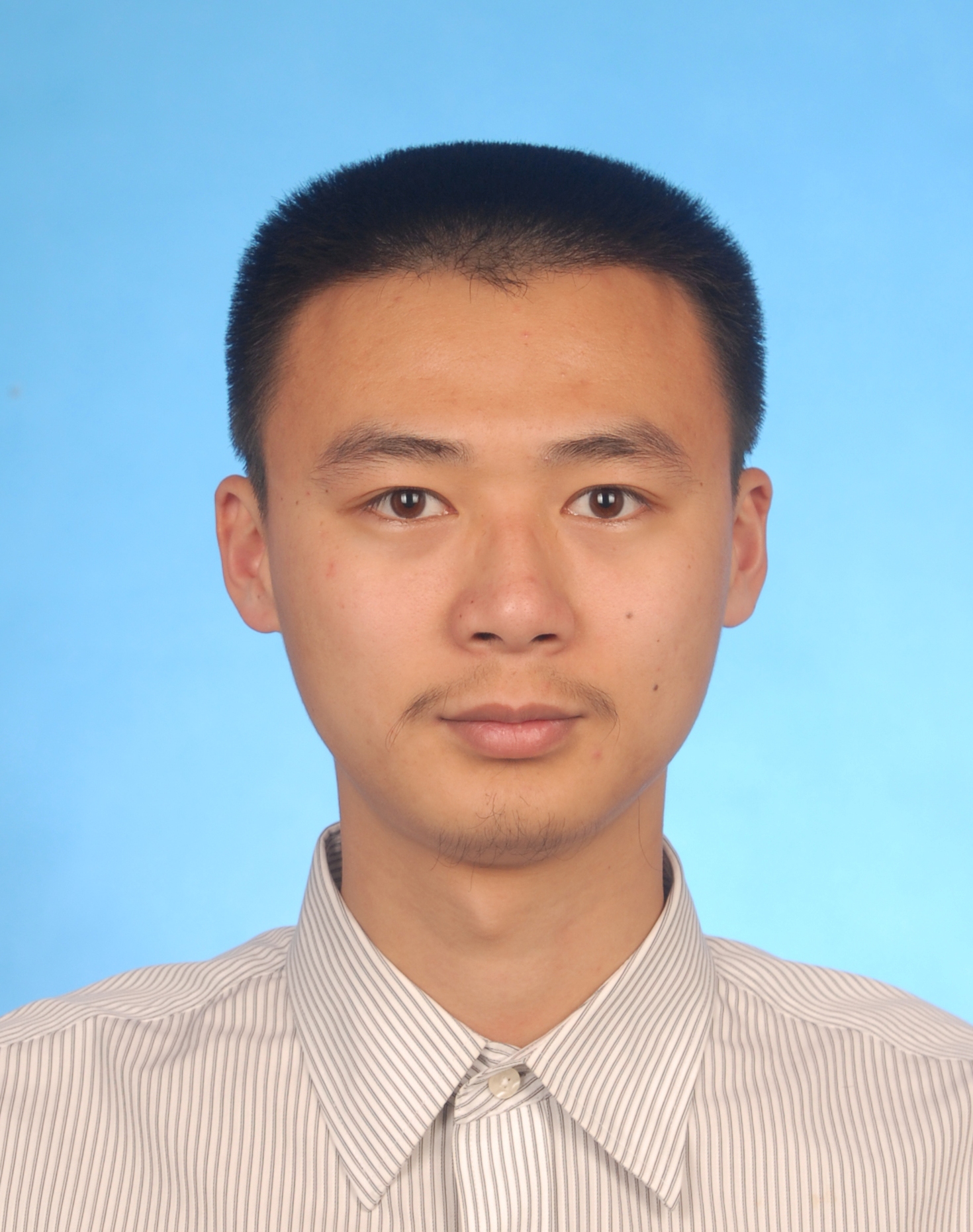}}]{Xiaoyu Du}
is currently a lecturer in the School of Software Engineering of Chengdu University of Information Technology, Chengdu, 
a visiting scholar in the NExT++ research center of National University of Singapore,
and a Ph.D. candidate of University of Electronic Science and Technology of China, Chengdu. 
He received his M.E. degree in computer software and theory in 2011 and B.S. degree in computer science and technology in 2008, both from Beijing Normal University, Beijing. 
His research interests include information retrieval, computer vision, and machine learning.
\end{IEEEbiography} \vspace{-20pt}

\begin{IEEEbiography}[{\includegraphics[width=1in,height=1.25in,clip,keepaspectratio]{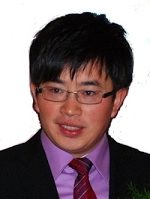}}]{Richang Hong}
Richang Hong received the PhD degree from
the University of Science and Technology of
China, Hefei, China, in 2008. He is a professor with the Hefei University of Technology, Hefei, China. He was a research fellow in the School of Computing, National University of Singapore,
from Sep. 2008 to Dec. 2010. He has coauthored more than 70 publications in the areas of his research interests, which include multimedia content analysis and social media. He received the Best Paper Award in the ACM Multimedia 2010, Best Paper Award in the ACM ICMR 2015, and the Honorable Mention of the IEEE Trans. on Multimedia Best Paper Award 2015. He served as the associate editor of the IEEE Multimedia Magazine, Information Sciences and Signal Processing, Elsevier and the technical program chair of the MMM 2016 and ACM ICIMCS 2017. He is a member of the IEEE, the ACM and the executive committee member of the ACM SIGMM China Chapter.
\end{IEEEbiography} \vspace{-20pt}

\begin{IEEEbiography}[{\includegraphics[width=1in,height=1.25in,clip,keepaspectratio]{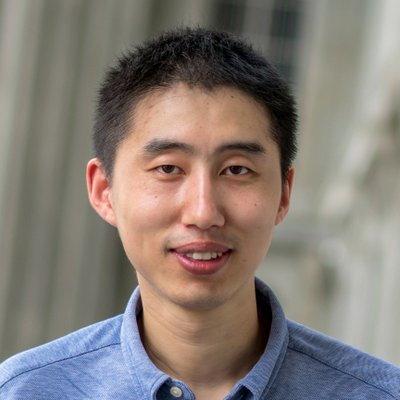}}]{Tongwei Ren}
Tongwei Ren is currently an associate professor with the State Key Laboratory for Novel Software Technology, Nanjing University. He received his Ph.D. in computer science and technology from Nanjing University. His research interest mainly includes multimedia computing and computer vision. He has over 50 publications appeared in international conferences such as MM, ICCV, and AAAI, and journals such as TIP and NEUCOM. His works on image analysis have received the Best Paper Award Honorable Mention of ICIMCS 2014 and the Best Paper Runner-up of PCM 2015. He has served as the workshop chair of ICIMCS 2015, the publication chair of PCM 2017 and the program chair of ICIMCS 2018. He has also served as the PC member for conferences including BigMM, ICIP, and PCM, and the regular reviewer for journals including TIP, TOMM and TMM.
\end{IEEEbiography} \vspace{-20pt}

\begin{IEEEbiography}[{\includegraphics[width=1in,height=1.25in,clip,keepaspectratio]{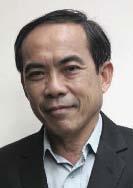}}]{Tat-Seng Chua}
is the KITHCT Chair Professor at the School of Computing, National University of Singapore. He was the Acting and Founding Dean of the School during 1998-2000. Dr Chuas main research interest is in multimedia information retrieval and social media analytics. In particular, his research focuses on the extraction, retrieval and question-answering (QA) of text and rich media arising from the Web and multiple social networks. He is the co-Director of NExT, a joint Center between NUS and Tsinghua University to develop technologies for live social media search. Dr Chua is the 2015 winner of the prestigious ACM SIGMM award for Outstanding Technical Contributions to Multimedia Computing, Communications and Applications. He is the Chair of steering committee of ACM International Conference on Multimedia Retrieval (ICMR) and Multimedia Modeling (MMM) conference series. Dr Chua is also the General Co-Chair of ACM Multimedia 2005, ACM CIVR (now ACM ICMR) 2005, ACM SIGIR 2008, and ACM Web Science 2015. He serves in the editorial boards of four international journals. Dr. Chua is the co-Founder of two technology startup companies in Singapore. He holds a PhD from the University of Leeds, UK.
\end{IEEEbiography} \vspace{-20pt}

\end{document}